\documentclass[11pt]{article}
\usepackage[margin=1in]{geometry}
\usepackage{times}
\usepackage[compact]{titlesec}
\usepackage{amsfonts}
\usepackage{amsmath}
\usepackage{amsthm}
\usepackage{amssymb}
\usepackage{latexsym}
\usepackage{epic}
\usepackage{epsfig}
\usepackage{hyperref}
\usepackage{verbatim}
\usepackage[justification=centering]{caption}
\usepackage{color}
\usepackage{xcolor}
\allowdisplaybreaks[1]
\usepackage{natbib}
\usepackage{algorithm}
\usepackage{algorithmic}
\usepackage{setspace}
\usepackage{derivative}
\usepackage{cancel}
\usepackage{array}
\usepackage{tabu}
\usepackage{pdfpages}
\usepackage{multicol}
\usepackage{graphicx}
\usepackage{subcaption}
\usepackage{mathtools}

\newcommand{\dd}{\mathrm{d}}

\newcommand{\seller}{the seller}
\newcommand{\buyer}{the buyer}
\newcommand{\Seller}{the seller}
\newcommand{\Buyer}{the buyer}

\newtheorem{theorem}{Theorem} 

\newtheorem{lemma}{Lemma}
\newtheorem*{lemma*}{Lemma}
\newtheorem{proposition}{Proposition}

\newtheorem{assumption}{Assumption}
\newtheorem*{proposition*}{Proposition}

\newtheorem{definition}{Definition}
\newtheorem{remark}{Remark}

\newtheorem*{conjecture*}{Conjecture}
\newtheoremstyle{nonindented}{1ex}{1ex}{}{}{\bfseries}{.}{.5em}{}
\newtheoremstyle{indented}{1ex}{1ex}{\itshape\addtolength{\leftskip}{0.6cm}\addtolength{\rightskip}{0.6cm}}{}{\bfseries}{.}{.5em}{}
\theoremstyle{nonindented}

\theoremstyle{indented}
\newtheorem*{direction*}{Research Direction}
\theoremstyle{plain}

\newcommand\blfootnote[1]{%
  \let\thefootnote\relax%
  \footnotetext{#1}%
  \let\thefootnote\svthefootnote%
}

\renewcommand{\bar}{\overline}

\def\min{\qopname\relax n{min}}
\def\max{\qopname\relax n{max}}

\def\argmax{\qopname\relax n{argmax}}

\def\Pr{\qopname\relax n{\mathbf{Pr}}}
\def\Ex{\qopname\relax n{\mathbf{E}}}

\newcommand{\RR}{\mathbb{R}}

\newcommand{\eat}[1]{}

\newcommand{\maxi}[1]{\mbox{maximize} & {#1 } & \\}

\newcommand{\st}{\mbox{subject to} }

\newcommand{\con}[1]{&#1 & \\}
\newcommand{\qcon}[2]{&#1, & \mbox{for } #2.  \\}
\newenvironment{lp}{\begin{equation}  \begin{array}{lll}}{\end{array}\end{equation} }
\newenvironment{lp*}{\begin{equation*}  \begin{array}{lll}}{\end{array}\end{equation*}}

    \title{Optimal Pricing of Information}
    
\author{Shuze Liu\\  University of Virginia \\ \small sl5nw@virginia.edu 
        \and 
        Weiran Shen\\ Renmin University of China  \\ \small shenweiran@ruc.edu.cn
        \and 
        Haifeng Xu\\  University of Virginia \\ \small hx4ad@virginia.edu }
\begin{document}
 
\begin{titlepage}

\date{}

\maketitle
\thispagestyle{empty}

\blfootnote{  We thank EC'2021 reviewers and audiences for helpful comments. Haifeng Xu would like to thank Dirk Bergmann,  Margaret Meyer, Yining Guo, and seminar participants of the 2021 Workshop on Strategic Communication and Learning for inspiring suggestions. 
    }

    \begin{abstract}
        
     A decision maker is deciding between an active action (e.g., purchase a house, invest certain stock) and a passive action.  The payoff of the active action depends on the buyer's private type and also an unknown \emph{state of nature}.  
     An information seller can design experiments to reveal information about the realized state to the decision maker, and  would like to maximize profit from selling such information.  We  fully characterize, in closed-form, the revenue-optimal information selling mechanism for the seller.
        After eliciting the buyer's type, the optimal mechanism charges the buyer an upfront payment and then simply reveals whether the realized state passed a certain threshold or not. The optimal mechanism features both price discrimination and information discrimination.  
       The special buyer type who is a priori indifferent between the active and passive action benefits the most from participating the mechanism. 
    \end{abstract}

       
\end{titlepage}

    
        
    

    



\section{Introduction}
\label{sec:intro}


















In numerous situations, a decision maker wishes to take an active move but is uncertain about its outcome and payoff. Such active moves range  from financial decisions of investing a stock or startup to daily-life decisions of purchasing a house or a used car, from macro-level enterprise decisions of developing a new product to micro-level decisions of approving a loan applicant or displaying online ads to a particular Internet user. In  all these situations, the decision maker's payoff for the active move relies on uncertain information regarding, e.g.,   potential of the invested company, quality of the house, popularity of the new product,  credit of the loan applicant, etc. Certainly,  the decision maker typically also has a passive backup option of not making the move, in which case he obtains a safe utility without any risk. To  decide between the \emph{active}   and  the \emph{passive} action, the decision maker can turn to an information seller who can access more accurate information about the uncertainties and thus help to better estimate the payoff for his action. Given the value of the seller's information to the decision maker, the seller can make a profit from how much the information  helped to  improve  utilities of the decision maker, i.e., the information buyer. 


This paper studies how a monopolistic information \emph{seller} (she)  can design an optimal pricing mechanism to sell her information to an information \emph{buyer} (he). The buyer (a decision maker) needs to take one of two actions. The \emph{active} action results in a payoff $v(q,t)$ where $t$ captures the buyer's private \emph{type} and the \emph{state of nature} $q$ summarizes the payoff-relevant uncertainty unknown to the buyer. The \emph{passive}   action for the buyer always results in the same utility, normalized to $0$, regardless of $q,t$. Both $q$ and $t$ are  random variables drawn independently from publicly known distributions. That is, the type $t$ captures the buyer's private preference and is assumed to be irrelevant to the informational variable $q$.\footnote{This independence assumption is relaxed in subsection \ref{section-dependence}.}   
The seller can design experiments to reveal partial information about    state $q$, and would like to design an optimal  mechanism to sell her information to a buyer  randomly drawn from the type distribution. We assume both the experiment itself and its outcomes (i.e., realized signals)  are contractible. 

As an example, consider a credit assessment company selling  credit information to a loan company.  In the loan company's payoff function $v(q,t)$ of the active action, informational variable $q$ captures the credit information of a randomly arriving loan applicant and can only be observed by the credit company. Type $t$ captures  the loan company's profit from the loan given that the applicant will pay back the loan on time, and is  independent of the applicant's credit information $q$.   While our model allows $q,t$  to be abstract variables from measurable sets  in general  (e.g., $q$ may contain employment history, loan history, etc.), it will be conceptually convenient to think of $q,t$ as numerical variables. For instance, consider $v(q,t) = qt- 2$ where: (1) $t$ is the loan company's profit from the loan; (2) $q\in[0,1]$ is a particular applicant's payback rate which can be estimated by the credit company through  data-driven prediction techniques today; (3) constant $2$ integrates operation  costs. The \emph{passive action} of rejecting  the loan applicant results in utility $0$.  We shall   capture the credit company's optimal mechanism for selling its payback rate prediction $q$.    


The  described problem setup above is a very basic monopoly pricing problem. However,  the problem of selling information turns out to differ significantly from  the classic pricing problem for selling goods. First, when selling (physical or digital) goods,  the seller's allocation rule can be described by a probability of giving out the goods and a risk-neutral buyer's utility is linear in the allocation variable. However, when revealing information to a buyer through experiments, the design variable of an experiment  for each buyer type  is high-dimensional or can even be a functional when the state is a continuum. Moreover, the buyer's utility is generally non-linear in the variables that describe an experiment \citep{bergemann2019information}.   Second,  in selling goods,  any individually rational buyer would participate   as long as their expected utility is at least $0$. However, in our setup of selling information, the buyer may already have  positive utility from his active action even without participating in the mechanism. An individually rational buyer would  participate in the mechanism only when his utility will become even higher. These differences make the seller's optimization task more challenging. This will be evident later in our characterization of the optimal mechanism, which turns out to be significantly different from, and arguably more intricate than,  the optimal pricing mechanism for selling goods by \cite{myerson81}.   

\subsection*{Main Result}
We consider the above information selling problem and characterize in closed-form the revenue-optimal mechanism, among all \emph{sequential  mechanisms} that includes all possible ways through which the seller may sequentially reveal information and ask for payments. 
To simplify the exposition, we assume that the buyer's value function is \emph{linear} and \emph{monotone non-decreasing} in $t$, i.e., $v(q,t) = \alpha(q)[t+\beta(q)]$ for some $\alpha(q) \geq 0$ and $\beta(q)$. In Subsection \ref{generalized-utility}, we discuss  how our  analysis and results can   be  generalized to  any  convex and monotone (in $t$) value functions $v(q,t)$.   

Assuming $v(q,t) = \alpha(q)[t+\beta(q)]$, we show that there always exists an optimal mechanism of a simple format --- a multi-entry menu where each entry containing  a threshold experiment and a payment for each buyer type.    
In this  optimal mechanism, the buyer is incentivized to report his true type $t$ first.\footnote{Equivalently, it is the best interest for each buyer type to choose the particular menu intended for him. That is,   the mechanism is incentive compatible. } The seller then charges the buyer $p_{t}$ and, afterwards, designs an experiment to reveal  whether the realized state $q$ satisfies $\beta(q) \geq \theta_t$ or not for some carefully chosen  threshold $\theta_t$. We thus call   the mechanism a \emph{threshold mechanism}. The thresholds and payments  generally vary for different buyer types, and are carefully designed to accommodate the amount of risk each buyer type can tolerate. That is, the optimal mechanism features both price discrimination and information discrimination.  We fully characterize the threshold and payment in the optimal mechanism. Depending on the setting,  the negative of the threshold (i.e., $-\theta_t$) turns out to equal either the (\emph{lower}) virtual value of type $t$ as defined by \cite{myerson81}, or its variant which we coin the \emph{upper} virtual value, or a novel convex combination of both coined the  \emph{mixed} virtual value.     

The above optimal mechanism exhibits multiple interesting properties. First,  the optimal mechanism turns out to only need to price the experiment with one round information revelation, even though the seller in our model is allowed to price experiment outcomes (i.e., signals) and use multiple rounds of information revelation. This is due to the independence of the informational variable $q$ and buyer type $t$, which makes an upfront payment and an ``aggregrated'' experiment without loss of generality. 
Second, the special buyer type $\bar{t}$ who is a-priori indifferent between active and passive action has the largest surplus from participating the mechanism. This is aligned with our intuition that this buyer type should benefit the most from additional information since the two actions appear indistinguishable to him a-priori.  Moreover, we show that the buyer surplus as a function of his type $t$ is increasing and convex when $t\leq \bar{t}$ but immediately transitions to be  decreasing and convex when $t \geq \bar{t}$. However, the buyer payment may be increasing or decreasing in $t$, depending on the setting. Third,  information discrimination turns out to be crucial for revenue. We show that if information discrimination is not allowed, i.e., suppose the same experiment must be used for all buyer types, then the best the seller can do in this case is to reveal full information and charge the Myerson's reserve price. We  demonstrate via an example that   the revenue in this case may be arbitrarily worse than the optimal.  However, under the monotone hazard rate assumption of the buyer type distribution, we show that the optimal single-entry menu can always guarantee at least $1/e(\approx 0.368)$ fraction of the optimal revenue.     

\subsection*{Related Works}

\textbf{Related works on selling information.}  The most related literature to our work is the recent study by \cite{bergemann2018design}, who also consider selling information to a decision maker. In their  model, the state of nature affects the payoff of every action. 
They characterize the optimal mechanism for the special cases with binary states and actions or with binary buyer types, whereas only  partial  properties about the optimal mechanism can be derived for the general case. In contrast, in our setup the state only affects the payoff of the buyer's active action. This restriction allows us to characterize the closed-form solution of the optimal mechanism with many (even continuous) states  and  buyer types, and for general buyer payoff functions. Moreover, our design space of mechanisms allows  multiple rounds of information revelation and also allows contracting the experiment outcomes (i.e., realized signals), though it turns out that  the optimal mechanism only needs to price  one-round experiments.\footnote{This is first observed by \cite{Babaioff12} yet we will provide a formal argument later for completeness.} While  \cite{bergemann2018design} also restrict their design space to mechanisms that only price one-round experiments, they pointed out that this restriction does lose generality in their general setup. That is,  the seller may derive strictly more revenue by using multi-rounds of experiments  or by contracting the experiment outcomes.

\cite{advice} studied the pricing of advice in a principal-agent model motivated by  consulting. The principal as a consultant in their model can contract the agent's actions. With such strong bargaining power, their main result shows that even the principal observes completely irrelevant information about the agent's payoffs, the principal can still obtain revenue that is as high as in the situation where she   fully observes the agents' payoffs. 
However, different from consulting service, our model of information selling assumes that only information itself (i.e., the experiment  or the experiment outcomes) is contractible and the buyer's actions are not contractible. Therefore, the main result of  \cite{advice} clearly does not hold in our model --- if the seller's information is irrelevant to the buyer's payoffs in our model, she will certainly get zero revenue. Interestingly,  the format of our optimal mechanism turns out to bear somewhat similar structure to the optimal contract of \cite{advice}, however our results are derived through different techniques and apply to much more general buyer value functions, whereas \cite{advice} restrict to   simpler agent utility functions (i.e., the sum of the agent type and the state) and only log-concave agent type distributions. 
\cite{horner2016selling} study the problem where a firm faces a decision on whether to hire an agent, who has a binary private type, i.e., competent or not. The firm  and agent can interact for many rounds by making money transfer and taking test to elicit information about the agent's type. They analyze the equilibrium when the number of rounds of interactions grows large. Both the  model and the nature of their results are different from us.  

There has also been recent interest of algorithmic studies that formulate optimization programs to compute the optimal mechanism for selling information to a decision maker.  \cite{Babaioff12} prove revelation principle types of results and characterize the format of the optimal mechanism, depending on whether the state and buyer type are correlated or not; they then 
develop optimization programs to compute the optimal mechanism. The efficiency for solving these programs were later improved by \cite{chen2020selling}.  

\textbf{Information revelation while selling goods.} Another relevant yet significantly different problem is the \emph{joint} design of information revelation scheme and the mechanism for selling goods, when the seller has private information about the goods.   
\cite{esHo2007optimal} studied revenue-maximizing mechanisms for selling a single indivisible good to multiple buyers when the auctioneer can  release, without observing, additional signals about the item.  
 \cite{reverse} derive closed-form optimal mechanism for selling goods to a single buyer with strategic disclosure of the seller's private information. In both models, it is still primarily the goods that are sold although, intuitively, part of their price includes the charge of the revealed information.  However, in our setting, the seller is a pure information seller without goods. This leads to significant technical differences in determining  participation constraints and how much information to reveal, and consequently leads to  different mechanism properties.  For instance, the payment function in their solution  is monotone decreasing in the buyer type whereas payment in our optimal mechanism may not even be monotone in the buyer type.     
 On the technical side, both works above rely on the Monotone Hazard Rate (MHR) assumption on buyer's type distribution whereas our results apply to general distributions. From the optimal mechanism design perspective,  \cite{10.1145/2940716.2940789} show that   the joint design of signaling schemes and auction mechanisms reduces to the design of multi-item auctions with additive buyer values.  \cite{bergemann2021selling} recently study information revelation in second-price auctions, motivated by the sale of impressions in online advertising.

\textbf{Contract design with outside options.}  
Our model is   conceptually related to contract design of countervailing incentives in principal-agent models with outside options \citep{countervailing1,countervailing3,countervailing2,countervailing4}. However, both the seller's objective and design space (e.g., information revelation schemes) in our model are significantly different. For instance, the principal's payoff depends on the agent's actions in these agency problems, whereas the seller's revenue only depends on the buyer's payment and nothing else.     
From a technical point of view, most related to us is the work of  \cite{countervailing2}. They systematically consider how  the function shape of the agent's outside option affects the agent's participation constraint, and consequently affects the format of the optimal mechanism. This is also one of the key technical challenges we had to address. However, 
the nature of their results   crucially differs from us ---  they cast the model as an optimal control problem and then analyze its properties, whereas we directly solve out from a closed-form optimal solution.   

\textbf{Information design. } 
Finally, our work is also relevant to the recent rich body of works on information design, a.k.a., bayesian persuasion \citep{kamenica2011bayesian}. 
Specifically, the most relevant literature to ours is the persuasion problem with a privately informed receiver \citep{kolotilin2017persuasion,guo2019interval}.  Similar to us, both papers study models with binary receiver actions. However, the design objective between persuasion and selling information is quite different and thus the solutions are not quite comparable.  Like us, \cite{kolotilin2017persuasion} also assume independence between the sender's information and receiver's private type, however the upper-censorship (or lower-censorship) structure of their optimal signaling scheme  differs from our threshold experiments.  \cite{guo2019interval} study persuasion when the receiver has private information about the state, captured as his   type.  They  show a nested interval   structure of the optimal signaling scheme, which is relevant to, yet still different from, our threshold  structure of the optimal disclosure.    

\section{Model and Problem Formulation}
\label{sec:model}



\subsection{The Setup}
We study the following  optimal information pricing problem  between an information \emph{seller} (she)   and an information \emph{buyer} (he).  The \buyer{} is a decision maker who faces one of two actions: a \emph{passive} action $0$ and an \emph{active}  action $1$.  The \buyer{} obtains an uncertain payoff $v(q,t)$ for the active action $1$  where   $q \in  Q$ is a random \emph{state of nature} unknown to the buyer and $t\in T$ is the buyer's private type. Both $T,Q$ are measurable sets.  The buyer's utility for the passive action $0$ is always $0$, irrespective of his type and the state of nature. In other words,  the passive action  is a   backup option for  the buyer.  
For example, if  the   \buyer{}  is a   potential purchaser of some goods (e.g., a house or a used car) with uncertain quality, the passive action $0$ corresponds to ``not purchase'' in which case   \buyer{} has no gain or loss,  whereas the active action $1$ corresponds  to ``purchase'' under which the \buyer's utility depends on the quality $q$ of the goods as well as how much he values the goods (captured by his private type $t$). 

Both $t$ and $q$ are  random variables that are  independently   distributed according to the cumulative distribution functions (CDF) $F(t)$ and $G(q)$, respectively.  We assume throughout the paper that that  both $F(t)$ and $G(q)$ are continuously differentiable, with corresponding probability density functions (PDF) $f(t)$ and $g(q)$.  
Both $F(t)$ and $G(q)$ are public knowledge. However, the realized  $q$ can only be   observed by the information seller. We study the \seller's problem of designing a \emph{revenue}-maximizing pricing mechanism  to sell her private observation of $q$ to  the  \buyer.  Notably,  \buyer{}'s private type $t$ is only known to   himself ---  
had  the \seller{} known   \buyer{} type $t$, the seller's optimal pricing mechanism is simply to  reveal full information   and then charge   \buyer{}  the \emph{value of (full) information} \citep{bergemann2018design}: $ \int_{q \in Q}  \max \{ 0, v(q,t) \}  g(q)\dd q  -  \max \{ 0, \int_{q \in Q}  v(q,t)  g(q) \dd q   \}$. 

Throughout we assume that the buyer payoff function $v(q, t)$ is monotone  non-decreasing  in  his  type $t$ for any $q \in Q$. For expositional simplicity,  we will  assume $v(q, t)$ is linear in $t$, i.e., there exist real-valued functions  $\alpha(q) \geq 0$ and $\beta(q)$ such that $v(q,t)=\alpha(q)(t+\beta(q))$. 
In Subsection \ref{generalized-utility}, we   show how our results and analysis easily generalize to any convex (in $t$) function $v(q, t)$. Linearity  also implies that the buyer's type $t \in \RR$ is a real value, which we assume is  supported on a closed interval $T = [t_1, t_2]$\footnote{ This implies that the type's density function $f(t)>0, \forall t\in T$.}. However, the state $q$  is allowed to be supported  on a general measurable set $Q$ and does  \emph{not} need  to be a real value. Such an abstract representation of $q$ is useful for accommodating applications where  $q$ may include the non-numerical features relevant to the \buyer's decisions (e.g., the brand and production time of a used car).  
Since $q$ is a random variable, $\beta(q)$ also has a probability distribution. For ease of presentation, we make a mild technical assumption  that the distribution of $\beta$ does not have any point mass.  
However, our analysis applies similarly to the general case in which $\beta(q)$ contains point masses, but just with more complex notations (see Appendix \ref{appendix:partial_recommendation} for more details).  

With slight abuse of notation, let $v(t)$ denote the buyer's expected utility for action $1$ under his prior beliefs about $q$, namely, when no information is purchased. That is,
\begin{equation}\label{def:buyer-expected-v}
    \text{Buyer's a priori utility of action 1: } \quad     v(t)=\int_{q \in Q} v(q,t) g(q) \dd q.  \quad 
\end{equation}

\subsection{Mechanism Space and the Revelation Principle} 
To maximize revenue,  the  \seller{} can design arbitrary mechanisms with possibly multiple rounds of interactions with the buyer. The task of designing a revenue-maximizing mechanism can be intractable unless a well-defined and general mechanism space is specified. Prior work  of \cite{bergemann2018design} restricts to the sale of experiments via only a single-round of information revelation. In this work, we consider a richer design space  of mechanisms, in which the  \seller{} is also allowed to contract the realized experiment outcomes (i.e., signals) and moreover, multiple rounds of information revelation and payments are allowed as well. Specifically, we consider the following set of \emph{sequential mechanisms}.\footnote{This general class of mechanisms was first introduced and studied by \cite{Babaioff12}, and was called the generic interactive protocols in their work. }   


\begin{definition}[\textbf{Sequential Mechanisms}]
A sequential mechanism is a mechanism that results in a finite extensive-form game between the seller and the buyer. Formally, let $C(n)$ be the set of all children nodes of node $n$. Then each non-leaf node $n$ in the game tree is one of the following three types:
\begin{itemize}
   \item \emph{Transfer node}, which is associated with a (possibly negative) monetary transfer $p(n)$ to the seller and has a single child node. 
   
    \item \emph{Seller node} that reveals information. Any seller node associates each state of nature $q$ with a distribution over $C(n)$, prescribing the probabilities of moving to its children nodes. That is, there is a function $\psi_n:Q\times C(n)\mapsto [0,1]$ for each seller node $n$ with $\sum_{c\in C(n)}\psi_n(c,q)=1, \forall q\in Q$. Thus,  a child node  $c$ carries information about $q$. 

    \item \emph{Buyer node}, which corresponds to an arbitrary set of   buyer choices with every choice leading to a child node.
\end{itemize}
\end{definition}
The buyer's final decision of taking the active or passive action is made \emph{after} the information selling process, and thus is not modelled in the above sequential mechanisms. Therefore, at each seller   node, the seller's   action is to choose a message to send to the buyer which determines the child node the game will move to; the buyer node has the similar functionality. Note that the mechanism is voluntary and the buyer is free to leave the mechanism at any stage.


When designing the revenue-optimal mechanism for selling physical goods, the celebrated revelation principle \citep{10.2307/1912346,10.2307/1914083} enables us to without loss of generality focus only on  truthful and  direct mechanisms. However, when selling information, sequential mechanisms can bring strictly more revenue than one-round mechanisms. We show that  our setting admits a stronger revelation principle  that allows us to consider w.l.o.g. the set of   truthful,  direct and one-round mechanisms. 

To describe the space of  \emph{one-round mechanisms}, we need the notion of \emph{experiments}, which formalize the way a \seller{} reveals information. Given a set of possible signals $\Sigma$, an experiment $\pi: Q \to \Delta_{\Sigma}$ is a mapping from the state    $q$ to a distribution over the signals in $\Sigma$. Such an experiment can be mathematically described by $\{ \pi(\sigma| q) \}_{q \in Q, \sigma \in\Sigma}$ where $\pi(\sigma|q)$ is the probability of sending signal $\sigma$ conditioned on state $q$. 
After observing signal $\sigma$,  the buyer infers posterior probability about any state $q$ via standard Bayes updates:

\begin{equation}
    g(q|\sigma) = \frac{\pi(\sigma|q) \cdot g(q)  }{ \int_{q'\in Q} \pi(\sigma|q') \cdot g(q') \dd q'}  = \frac{\pi(\sigma|q) \cdot g(q)  }{ \Ex_{q' \sim G} [\pi(\sigma|q')]   }.
\end{equation}
Consequently, conditioned on signal $\sigma$, if a buyer of type $t$ takes the active action, his expected utility is $\int_{q \in Q} v(q,t) g(q|\sigma) \dd q $. 

Different experiments reveal different amount of information to the buyer, and thus are of different values. A \emph{one-round mechanism} is a menu  of experiments and prices that results in a single-round of interaction between the seller and the buyer.

\begin{definition}[\textbf{One-round Mechanisms}]
A one-round mechanism $\mathcal{M}$, described by a  menu $\{ (p_t, \pi_t) \}_{t \in T}$,  proceeds as follows:
\begin{enumerate}
    \item The buyer is asked to report (possibly untruthfully) his type $t$;
    \item The seller charges the buyer $p_t$;
    \item The seller reveals information about $q$ according to experiment $\pi_{t}$.
\end{enumerate}
\end{definition}
A one-round mechanism can clearly be represented as a special sequential mechanism, with the 3 steps corresponding to a buyer node, followed by a transfer node, and then followed by a seller node. Though sequential mechanisms can generally contract experiment outcomes (when a seller node is followed by transfer nodes), any one-round mechanism only prices the experiment $\pi_t$ at price $p_t$  but does not contract the experiment outcomes.   

Let $U(t';t)$ denote the expected utility of a buyer with type $t$ reporting type $t'$, defined as
\begin{gather*}
U(t';t)=\sum_{\sigma\in \Sigma}\max\bigg\{ \int_{q\in Q}v(q,t)\pi_{t'}(\sigma|q)g(q)\,\dd q \, \, , \, \, 0 \bigg\}-p_{t'}.  
\end{gather*} 
A one-round mechanism is said to be \emph{incentive compatible}, if it is the buyer's best interest to report his type truthfully, i.e., $U(t;t)\ge U(t';t), \forall t, t' \in T$. The following revelation principle shows that it is without loss of generality to consider direct, incentive compatible mechanisms and one-round in our model. 

\begin{lemma}[\textbf{Revelation Principle}]    \label{lem:revelation}
For any sequential mechanism $\mathcal{M}$, there exists a direct, incentive compatible and one-round mechanism  that achieves the same expected revenue as $\mathcal{M}$.  
\end{lemma}
Standard revelation principle argument implies that the seller can w.l.o.g incentivize truthful type report at the beginning. To prove   Lemma \ref{lem:revelation}, the non-trivial part is to  argue that a single-round of payment and information revelation suffice. This is a consequence of our independence assumption between state $q$ and buyer type $t$, such that it allows us to simply combine all steps of information revelation as a single experiment and combine all payments as a single upfront payment. A formal proof is deferred to Appendix \ref{append:revelation}.  Notably, the proof of Lemma \ref{lem:revelation} relies crucially on the independence of state $q$ and buyer type $t$. Fundamentally, this is because with correlation among the buyer type and state, a buyer type $t$, if misreporting $t'$, will perceive a different expected payment  as the $p_{t'}$ perceived by the buyer type $t'$ since $t$ and $t'$ hold different belief about $q$ and thus the expected payments w.r.t. each signal realization (see the proof for more illustration). 

Next, we further simplify the mechanism design space. First, we show in Lemma \ref{lem:positive-pay} that it is  without loss of generality to consider mechanisms  with non-negative payments. 
While this result is intuitive, we point out that it does not trivially hold. In fact, when $q$ and $t$ are correlated, the full-surplus-extracting sequential mechanism of \citep{Babaioff12} may   have to use \emph{negative} payments.  The proof of this lemma is deferred to Appendix \ref{appendix:positive-pay}. 
\begin{lemma}[\textbf{Non-Negative Payments}]
    There exists an optimal IC, IR  and one-round mechanism in which $p_t \geq 0$ for all $t \in T$. 
    \label{lem:positive-pay}
\end{lemma}

Second, the following known result of \cite{bergemann2018design} shows that when pricing experiments, we can without loss of generality price \emph{responsive} experiments, in which each signal  leads to a unique buyer best response action.
From this perspective, each signal in a responsive experiment can be viewed as an \emph{obedient} action recommendation. 

\begin{lemma}[\cite{bergemann2018design}]
The outcome of any mechanism can be obtained by using responsive experiments. 
\label{lem:signal-space}
\end{lemma} 

\subsection{Formulating the Optimal Pricing Problem} 
Based on the above simplification of the design space, we now formulate the mechanism design problem.  We start by introducing (functional) variables to describe a one-round mechanism with responsive experiments. We will think of the payment in the menu $\mathcal{M}$ as a function  $p(t)$ of buyer types $t$. Since the buyer has two possible actions, any responsive experiment $\pi_t$ for buyer type $t$ only needs two signals. With slight abuse of notation, we use function $\pi(q,t)\in [0,1]$ to denote the probability of sending signal \texttt{active} (interpreted as an obedient   recommendation of the active action), conditioned on state realization $q$. Naturally, $[1- \pi(q,t)]$ is   the probability of sending signal \texttt{passive}   conditioned on state $q$.   Our  goal is to derive a feasible menu ---  represented by functions $\pi^*(q, t)$ and $p^*(t)$ ---   that maximizes the seller's revenue. 
\begin{gather*}
   \text{Seller Revenue: } \quad   \max_{\pi, p} \int_{t\in T} f(t) p(t) \,\dd t.
\end{gather*}
Note that this is  a \emph{functional optimization} problem since both $\pi(q,t), p(t)$ are  functional variables that depend on continuous variable $t \in [t_1, t_2](=T)$ and abstract variable $q$ from a measurable set $Q$.  
The remainder of this section is devoted to formulating constraints on $\pi(q,t), p(t)$  according to Lemma \ref{lem:revelation},  \ref{lem:positive-pay} and  \ref{lem:signal-space}. 

\vspace{2mm}
\noindent {\bf Obedience constraints.} Lemma \ref{lem:signal-space} shows that any responsive experiment only needs to have two signals which make obedient recommendation of the active and   passive action, respectively. This poses two constraints on the   function $\pi(q,t)$:(1) $ 
    \int_{q \in Q} \pi(q, t) v(q,t) g(q) \,\dd q \geq   0,   \forall t \in T$; (2) $  
    \int_{q \in Q} [1-\pi(q, t)] v(q,t) g(q) \,\dd q \\ \leq   0,   \forall t \in T$.  
The first constraint above ensures that when signal \texttt{active} is sent to   buyer  type $t$, the buyer's expected value $ \frac{ \Ex_{q\sim G}\left[ \pi(q,t)v(q,t) \right] }{ \Ex_{q\sim G} [\pi(q,t)] }$ for taking the active action is indeed at least $0$, which is the expected value of taking the passive action. Similarly, the second constraint ensures the obedience of the \texttt{passive} signal. 
Slightly manipulating the second constraint above, we obtain  $\int_{q \in Q}  \pi(q, t)  v(q,t) g(q) \,\dd q \geq   \int_{q \in Q}  v(q,t) g(q) \, \\ \dd q  = v(t)$, where $v(t)$ defined in Equation \eqref{def:buyer-expected-v} is the buyer's a priori expected value of the active  action.  Therefore, we can conveniently summarize the obedience constraint as follows: 
\begin{gather}
    \text{Obedience: } \quad \int_{q \in Q} \pi(q, t) v(q,t) g(q) \,\dd q \geq   \max \{ 0, v(t) \},  \forall t \in T. \label{cons:obedience}
\end{gather} 
\vspace{2mm}
\noindent {\bf Individual rationality (IR) constraints.} 
Since the \buyer{} gets utility $0$ from the passive action,  the expected   utility of buyer type $t$, if he reports his type \emph{truthfully} and follows the seller's obedient recommendation, is  
\begin{equation} \label{def-u(t)}
        u(t)=\Ex_{q \sim G}[\pi(q,t)v(q,t)]-p(t)=\int_{q\in Q}\pi(q, t)v(q,t)g(q)\,\dd q -p(t),
\end{equation}
where the first term is the value   from his decision making assisted by the seller's information  and the second term is the payment to the seller. 
To ensure the buyer's participation in the mechanism,   the following individual rationality (IR) constraint is required:
\begin{gather}
    \text{IR:} \quad \int_{q \in Q} \pi(q, t) v(q,t) g(q) \,\dd q - p(t) \geq  \max \{ 0, v(t) \}, \forall t \in T, \label{cons:IR}
 \end{gather}
where the right-hand side is the buyer's  expected utility of not participating in the mechanism and simply takes the  best action according to his prior belief about $q$. Interestingly, since the payment function is always non-negative according to Lemma \ref{lem:positive-pay}, the IR constraint \eqref{cons:IR} turns out to imply the obedience constraint \eqref{cons:obedience}.   

The buyer surplus   $s(t)$   --- the additional utility gain of participating in the mechanism ---  as a function of the buyer type $t$ is defined as follows:  
\begin{equation} \label{eq:buyer-surplus}
  \text{Buyer surplus:}  \quad  s(t)= \int_{q \in Q} \pi(q, t) v(q,t) g(q) \,\dd q - p(t)-\max\{v(t),0\}.
\end{equation} 
The IR Constraint \eqref{cons:IR} is equivalent to non-negative surplus.

\vspace{2mm}
\noindent {\bf Incentive compatibility (IC) constraints.} The derivation of the IC constraints turns out to be more involved.  IC requires that when reporting truthfully, a buyer of type $t$ should obtain a higher utility than misreporting any other type $t'$. This turns out to require some analyses since when a buyer of type $t$ misreports type $t'$, the resulting experiment $\{ \pi(q,t') \}_{q \in Q}$ may not be obedient for $t$ any more, leading to non-linearity in  the IC constraints. Specifically, upon receiving signal \texttt{active},   the expected value of the active action for a type-$t$ buyer misreporting $t'$  is
\begin{equation}\label{eq:IC-derivation1}
    V_a(t';t)  \vcentcolon = \int_{q \in Q} \pi(q, t') v (q,t)   g(q) \,\dd q =   \int_{q \in Q} \pi(q, t') \alpha(q)[t+\beta(q)] g(q) \,\dd q. 
\end{equation} 
Since $\pi(q,t')$ may not be obedient for buyer type $t$, he will choose between active action and the passive action, leading to true expected value  $\max\{V_a(t';t), 0\}$ in this situation.  

Similarly, upon receiving signal \texttt{passive}, the buyer's value is the maximum between $0$ and the following:
\begin{equation}\label{eq:IC-derivation2}
    \int_{q \in Q} [1-\pi(q, t')] v(q,t)  g(q) \,\dd q = v(t) - V_a(t';t).
\end{equation}  
Combining both situations, the expected utility obtained by a buyer of type $t$ from misreporting type $t'$ is  $\max\{V_a(t';t), 0 \} + \max \{v(t) - V_a(t';t), 0 \}-p(t')$. So the incentive compatibility constraint becomes the following: 
 \begin{gather}\label{eq:IC-original}
 u(t) \geq \max\{V_a(t';t), 0 \} + \max \{v(t) - V_a(t';t), 0 \}   - p(t').
\end{gather} 
Such non-linear constraints are  difficult to handle in general. Interestingly, it turns out that we can leverage previous results to  reduce Constraint \eqref{eq:IC-original} to linear constraints on $\pi$, with some careful case analysis:
\begin{enumerate}
    \item When $t >  t'$, we have $  V_a(t';t) \ge V_a(t';t') \ge 0$, where the first inequality is due to the assumption $\alpha(q)\ge 0$ and the second comes from the obedience constraint \eqref{cons:obedience} for $t'$. In this case, the right-hand side of  Constraint \eqref{eq:IC-original} becomes $V_a(t';t) + \max \{v(t) - V_a(t';t), 0 \}   - p(t')$, or equivalently $\max \{v(t),   V_a(t';t) \}   - p(t')$. Note that $u(t) \geq v(t) - p(t')$ is already implied by the IR constraint $u(t) \geq v(t)$ and the condition $p(t')\geq 0$. Therefore, the only non-redundant constraint in this case is $u(t) \geq V_a(t';t) - p(t')$. 
    
    \item When   $t <   t'$, we have $  v(t) - V_a(t';t)\le v(t') -  V_a(t';t') \le 0 $ for similar reasons. In this case, the right-hand side of the above constraint becomes $\max \{  V_a(t';t), 0 \}    - p(t')$. Again, $u(t) \geq   - p(t')$ is already implied by the IR constraint $u(t) \geq 0$  and the condition $p(t')\geq 0$.  Therefore, the only non-redundant constraint in this case is also  $u(t) \geq V_a(t';t) - p(t')$. 
\end{enumerate}  
To summarize, given the IR and non-negative payment constraints, the IC constraint can finally be reduced to  the following: 
\begin{gather}
    \text{IC:} \quad \int_{q \in Q} \pi(q, t) v(q,t) g(q) \,\dd q - p(t) \geq \int_{q \in Q} \pi(q, t') v(q,t) g(q) \,\dd q  - p(t'), \forall t, t' \in T. \label{cons:IC}
\end{gather}

\vspace{2mm}
\noindent {\bf Combined optimization problem.} The   derivation and simplification above ultimately lead to the following optimization problem, with functional variables $ \pi(q,t), p(t)$: 
\begin{lp}\label{lp:opt-formulation}
\maxi{ \int_{t\in T} f(t) p(t) \,\dd t.}
\st 
\qcon{\int_{q \in Q} \pi(q, t) v(q,t) g(q) \,\dd q - p(t) \geq  \max \{ 0, v(t) \} }{ t \in T}
\qcon{\int_{q \in Q} [\pi(q, t) - \pi(q,t')] v(q,t) g(q) \,\dd q - p(t) + p(t') \geq 0}{t, t' \in T}
\con{p(t) \geq 0,  \quad \pi(q,t) \in [0,1]}  
\end{lp}

\section{The Optimal Mechanism}

In this section, we present  the characterization of the optimal pricing mechanism. Mathematically, we derive an optimal solution in closed-form to the functional optimization problem   \eqref{lp:opt-formulation}.   The optimal mechanism we obtain turns out to belong to the following category of \emph{threshold mechanisms}.
\begin{definition}[\textbf{Threshold Mechanisms}]
    A mechanism  $(\pi, p)$ is called a threshold mechanism if  it only uses threshold experiments. That is, there exists a   function $\theta(t)$, such that for any $t \in [t_1, t_2]$, 
    \begin{gather*}
        \pi(q, t)=
        \begin{cases}
            1 & \mathrm{if~} \beta(q)\ge \theta(t)\\
            0 & \mathrm{otherwise}
        \end{cases}.
    \end{gather*}
In this case, $\pi(q,t)$ is fully described  by the \emph{threshold function}  $\theta(t)$. 
\label{def:threshold}
\end{definition}
Note that the term ``threshold'' is only a property about the experiments and does not pose any constraint on the payment function $p(t)$.  To formally present our mechanism, we will need the following notions of \emph{lower}, \emph{upper} and \emph{mixed} virtual value functions. 
\begin{definition}[\textbf{Lower/Upper/Mixed Virtual Value function}]
   For any type $t$ with PDF $f(t)$ and  CDF $F(t)$, the function $\underline{\phi}(t) = t - \frac{1-F(t)}{f(t)}$ is called  the \emph{lower virtual value function} and $\bar{\phi}(t) = t + \frac{F(t)}{f(t)}$ is called the \emph{upper virtual value function}. Moreover, for any $c\in [0,1]$,  $ \phi_c(t) = c\underline{\phi}(t) + (1-c) \bar{\phi}(t)$ is called a \emph{mixed virtual value function}. 
   
   Any virtual value function  is \emph{regular} if it is monotone non-decreasing in $t$.
\end{definition}
The lower virtual value function $\underline{\phi}(t)$ is precisely the virtual value function   commonly used in  classic mechanism design settings \citep{myerson81}. We remark that while the upper and mixed virtual value function were not formally defined before, they have implicitly shown up in previous works  and typically give rise when the IR constraints are binding at the largest type (e.g., \cite{advice}). However, the specific formulation for the information selling problem allows us to characterize the optimal mechanism for much more general buyer utility functions (see more detailed comparison in the related work). 



\textbf{Ironing.} When a virtual value function is irregular, we will need to apply the so-called ``ironing'' trick to make it monotone non-decreasing in $t$.  \cite{myerson81}  developed a   procedure for ironing the lower virtual value function $\underline{\phi}(t)$. This procedure can be easily generalized to iron any function about the buyer type $t$, specifically, also  to the three types of  the virtual value functions defined above. For any virtual value function $\phi(t)$ (upper, lower or mixed),  let $ \phi ^{+}(t)$  denote  the ironed version of  $\phi(t)$ obtained via the standard ironing procedure of \cite{myerson81} (for completeness,  we give  a formal description of this ironing   procedure in  Appendix \ref{append:ironing-procedure}).\footnote{For techniques to iron a general function, we refer the reader to a recent work by \cite{toikka2011ironing}.}


If a virtual value function $ \phi (t)  $ is already   non-decreasing,  it   remains the same after   ironing, i.e., $\phi^+(t) = \phi(t), \forall t$. With  ${\phi}_c(t) = c\underline{\phi}(t) + (1-c) \bar{\phi}(t)$, the following useful properties of the ironed mixed virtual value functions will be needed for proving our main result (and may also be of independent interest in general). Their proofs are technical and are deferred to Appendix \ref{appendix:virtual_value_order_proof}. 

\begin{lemma}
[\textbf{Useful Properties of Ironed Mixed Virtual Values}]  

\hspace{22mm}
 \begin{enumerate}
     \item For any $0 \leq c < c' \leq 1$,  ${\phi}_c^+ (t) \geq {\phi}_{c'}^+(t)$ for any $t$;  
     \item For any $c \in [0 ,1]$, let  $t_c$ be the buyer type such that $F(t_c)  = c$. Then we have $\phi_c^+(t)\leq t, \forall t\leq t_c$ and $\phi_c^+(t)\geq t, \forall t\geq t_c$. This also implies $ \underline{\phi}^{+}(t)< t< \bar{\phi}^{+}(t) , \forall t \in (t_1, t_2)$.
 \end{enumerate}
\label{lem:virtual_value_order}
\end{lemma}

Notably, the second property above also implies that $\phi_c^+(t_c) = t_c$ always holds.    


We will be readily prepared to state the optimal mechanism after introducing the following two quantities:
  \begin{eqnarray}\label{eq:upper-lower-bound1} 
        V_L&=&\max \{v(t_1), 0 \} + \int_{t_1}^{t_2} \int_{q: \beta(q) \geq - \underline{\phi}^+(x)} g(q) \alpha(q) \,\dd q\dd x, \\ \label{eq:upper-lower-bound2}  
        V_H &=&\max \{v(t_1), 0 \} + \int_{t_1}^{t_2} \int_{q: \beta(q) \geq - \bar{\phi}^+(x)} g(q) \alpha(q) \,\dd q\dd x,
    \end{eqnarray}
where $\bar{\phi}^+(x)$ and $\underline{\phi}^+(x)$ are the ironed upper and lower virtual value functions, respectively. Note that Lemma  \ref{lem:virtual_value_order} implies $- \underline{\phi}^+(x) \geq - \bar{\phi}^+(x)$ and consequently $V_L \leq V_H$ since $g(q) \alpha(q)$ is always non-negative and thus $V_L$ integrates over a smaller region.  

Our main result  is then summarized in the following theorem. 

\begin{theorem}[\textbf{Characterization of an Optimal Mechanism}]
\quad
\begin{enumerate}
\item If $v(t_2) \leq V_L $, the threshold mechanism with  threshold   function $\theta^*(t) =  -\underline{\phi}^+(t) $ and the following payment function represents an optimal mechanism:
\begin{gather*}
    p^*(t) = \int_{q\in Q}  \pi^*(q,t)g(q)  v(q,t) \,\dd q -  \int_{t_1}^{t} \int_{q \in Q}  \pi^*(q, x) g(q) \alpha(q) \,\dd q\dd x.
\end{gather*}
where $\pi^*$ is determined by $\theta^*(t)$ as in Definition \ref{def:threshold}. 
Moreover, $p^*(t)$ is monotone non-decreasing for $t\in[t_1, t_2]$. 
\item 
If $v(t_2) \geq V_H $,   the threshold mechanism with  threshold 
function $\theta^*(t) =  -\bar{\phi}^+(t) $ and the following payment function represents an optimal mechanism:
\begin{gather*}
    p^*(t) = \int_{q\in Q}  \pi^*(q,t)g(q)  v(q,t) \,\dd q +  \int_{t}^{t_2} \int_{q \in Q}  \pi^*(q, x) g(q) \alpha(q) \,\dd q\dd x - v(t_2),
\end{gather*}
where $\pi^*$ is determined by $\theta^*(t)$ as in Definition \ref{def:threshold}. 
Moreover, $p^*(t)$ is monotone non-increasing for $t\in[t_1, t_2]$. 
\item If $V_L < v(t_2) < V_H$, let $c \in (0,1) $ be a constant that satisfies
 \begin{gather*}
    \int_{t_1}^{t_2}\int_{q:\beta(q)\ge-\phi_c^+(t)} g(q)  \alpha(q)\,\dd q \dd t = v(t_2),
 \end{gather*}
where $\phi_c^+(t)$ is the ironed version of the mixed virtual value function $\phi_c(t)$. 
Then the  threshold mechanism  with threshold function  $\theta^*(t)   =-\phi_c^+(t)$ and the following payment function represents an optimal mechanism:
\begin{gather*}
    p^*(t) = \int_{q\in Q}  \pi^*(q,t)g(q)  v(q,t) \,\dd q -  \int_{t_1}^{t} \int_{q \in Q}  \pi^*(q, x) g(q) \alpha(q) \,\dd q\dd x.
\end{gather*}        
Moreover,  $p^*(t)$ is monotone non-decreasing in $t$ when $F(t)\leq c$ and monotone non-increasing when $F(t)>c$. 
\end{enumerate}
Let $\bar{t}$ satisfy $v(\bar{t}) = 0$. In all cases above, the buyer surplus function $s(t)$ is convex and monotone non-decreasing when $t \leq \bar{t}$, but immediately transitions to be convex and monotone non-increasing when $t \geq \bar{t}$. 
\label{thm:opt-scheme}
\end{theorem}

The following are a few remarks  regarding Theorem \ref{thm:opt-scheme}.  

\begin{remark}
The optimal mechanism  generally features both price discrimination and information discrimination (see also a concrete example in Section \ref{susec:ex1}). This crucially differs from the sale of a physical goods to a buyer in which the optimal mechanism does \emph{not} exhibit price discrimination.  Notably, the price discrimination is a consequence of information discrimination. That is, for any two buyer types  $t, t'$, if their experiments are the same, then their payment must also be the same, i.e., $p^*(t) = p^*(t')$. This is a simple consequence of the IC constraint --- if  $p^*(t) > p^*(t')$, the buyer of type $t$ would misreport $t'$ by which she gets the same information but pays less. 
\end{remark}

\begin{remark} In all three cases of Theorem \ref{thm:opt-scheme}, a threshold mechanism is optimal, though the format of the optimal mechanism depends on how $v(t_2)$ compares to $V_L, V_H$.  Threshold mechanisms are ubiquitous in reality. In various formats of quality testing, inspection and recommendation services, we often pay for these ``experiments'' in order to see whether some goods pass  a test or or services  deserve  a recommendation. These can be viewed as threshold mechanisms for selling information. 
From this perspective, Theorem \ref{thm:opt-scheme} characterizes the optimal design for selling such experiments/information.     
 \end{remark}






\begin{remark}
We   briefly discuss the choice of the constant $c$ in Case 3 of Theorem \ref{thm:opt-scheme}. As we will show later in our proof, $v(t_2) \leq V_H$   implies $v(t_1) \leq 0$ for any feasible mechanism. Therefore, in Case 3, the $V_L, V_H$ defined in Equation \eqref{eq:upper-lower-bound1} and \eqref{eq:upper-lower-bound2}  only has the integral term. Therefore, the condition of Case 3 boils down to  $$\int_{t_1}^{t_2} \int_{q: \beta(q) \geq - \underline{\phi}^+(x)} g(q) \alpha(q) \,\dd q\dd x < v(t_2)  <  \int_{t_1}^{t_2} \int_{q: \beta(q) \geq - \bar{\phi}^+(x)} g(q) \alpha(q) \,\dd q\dd x.  $$ 
Since $\underline{\phi}^+(x) < \bar{\phi}^+(x)$ for any $x$, any $c \in (0,1)$ will ``interpolate'' the two integral region $\{q: \beta(q) \geq - \bar{\phi}^+(x)\}$ and $\{q: \beta(q) \geq - \underline{\phi}^+(x)\}$.  Since we assume that the distribution has no point mass, the following expression
\begin{gather*}
    \int_{t_1}^{t_2} \int_{q: \beta(q) \geq - \phi_c^+(x)} g(q) \alpha(q) \,\dd q\dd x
\end{gather*}
is continuous in $c$.\footnote{This is the only place where the assumption that the distribution of $\beta$ has no point masses is needed. Without this assumption, the threshold mechanism   will need randomization for those $q$ with $\beta(q) = \phi_c^+(t)$. See Appendix \ref{appendix:partial_recommendation} for the refined  characterization of the optimal mechanism for general $\beta$.  } 
Lemma \ref{lem:virtual_value_order} implies that it is also monotone weakly decreasing in $c$.
This thus leads to a unique choice  of the constant $c\in (0,1)$ that makes   the above term equal $v(t_2)$. We can pin down this $c$ via a simple binary search. 
\end{remark} 


\subsection{ An Example }\label{susec:ex1}
Consider the sale of credit assessment example in Section \ref{sec:intro} with $v(q,t) = qt - 2 = \alpha(q)[t+\beta(q)]$ where $\alpha(q) = q$ and $\beta(q) = \frac{-2}{q}$. Suppose $q \in Q = [0,1]$ is uniformly distributed, i.e., $g(q) = 1$. Let $t \in T = [2,3]$ also be uniformly distributed with $f(t) = 1$.\footnote{ Besides credit assessment, this setup also captures other applications such as online advertising. Here, $q$ is the probability that an  Internet user will purchase the product of an advertiser (the information buyer) and $t$ is the advertiser's revenue from selling a product. The constant $2$ captures the advertiser's payment  for displaying his ads to an Internet user. The information seller may be a  marketing company who can predict each Internet user's probability of conversion with her rich data and powerful machine learning technology. }  




In this example, $\underline{\phi}(t)$ is already non-decreasing, and thus the ironing procedure is not needed. We have $\underline{\phi}^+(t) = \underline{\phi}(t) = t - \frac{1 - F(t)}{f(t)} = 2t - 3$. Note that $v(t)  = \int_{q \in Q}g(q)v(q,t)\,\dd q = \int_{0}^1 (tq - 2)\, \dd q = \frac{t}{2}-2 $ for any $t \in [2,3]$.   Since $V_L$ defined in Equation \eqref{eq:upper-lower-bound2} is clearly non-negative, we have $v(t_2) = -0.5 < 0 \leq V_L$, so the instance falls into Case 1 of Theorem \ref{thm:opt-scheme}.  This implies that an optimal mechanism can be specified by a threshold experiment $\theta^*(t) = -\underline{\phi}^+(t) = 3 - 2t$. That is, for any buyer type $t$ the mechanism will make obedient recommendation of the active action when $\beta(q) \geq -\underline{\phi}^+(t)$,  or concretely,  when $q \geq \frac{2}{2t-3}$. Now there are two situations.
\begin{itemize}
    \item When $t < 2.5$, we have $\frac{2}{2t-3} > 1$. This means the mechanism will never recommend the active action since $q$ is at most 1. Therefore, we have $\pi^*(t,q) = 0$ for all $q\in Q$ in this case and the payment is $p^*=0$. For these buyer types, the seller simply sells no information to them and charges them $0$ as well. 
    \item When $t \ge 2.5$, the mechanism will recommend the active action when $q \geq \frac{2}{2t-3}$, which is a threshold in $(0,1)$ and decreases in $t$.  In this situation, the payment function $p^*(t)$ can then be computed as follows 
\begin{align*}
p^*(t) = \int _{\frac{2}{2t-3}}^1(qt\:-\:2)\,\dd q-\int _{2.5}^t\int _{\frac{2}{2x-3}}^1 q\,\dd q\dd x  = -0.25 +\frac{4t-9}{\left(2t-3\right)^2}. 
\end{align*}
For these buyer types, their utility from the mechanism will be  
\begin{align*}
u(t) =  \int _{\frac{2}{2t-3}}^1(qt\:-\:2)\,\dd q- p^*(t) 
= -1.75+\frac{t}{2}+\frac{1}{2t-3}.
\end{align*} 
\end{itemize}

Notably, to achieve the optimal revenue, the above mechanism does not simply recommend the active action whenever $v(q,t) \geq 0$.  For example, when   $t = 2.3$, the mechanism reveals no information (and asks for no charge as well) even  for $q$ with   $v(q,t) > 0$. Therefore, the revenue-optimal mechanism  generally uses non-trivial information structures.  
Moreover, the optimal mechanism uses a menu with infinitely many entries.    

\subsection{The Power of Information Discrimination}
The above example shows that the optimal mechanism features information discrimination, i.e.,  reveals different information to different buyer types, which then leads to price discrimination.  One might wonder how well a mechanism can perform if information discrimination is not allowed. Our following proposition shows that in this case,  the optimal mechanism is to simply post a uniform price and then reveal full information to any buyer who is willing  to pay. 

To describe the mechanism, we introduce a notation $e(t)$ that captures the value of full information for any buyer with type $t$: 
\begin{align} \label{e(t)}
e(t) = \int_{q\in Q} \max \{ v(q,t) \,,\, 0\}g(q)\, \dd q -    \max \bigg\{\int_{q\in Q}  v(q,t) g(q)  \,\dd q \, ,\, 0\bigg\} 
\end{align}
That is, $e(t)$ equals the additional value buyer type $t$ obtains by fully observing $q$. We then have the following proposition. 












\begin{proposition} \label{lem:full-revealing}
  If information discrimination is not allowed, then the optimal mechanism is to charge the Myerson's reserve price $r^*$ with respect to value $e(t)$, i.e., $r^* = \argmax_{r} \,  [ r \cdot  \Pr_{t\sim F} (e(t) \geq r) ] $, and then reveal full information to any buyer who pays.
\end{proposition}

The proof of Proposition \ref{lem:full-revealing} is straightforward. In any incentive-compatible   optimal  mechanism with a single experiment, the buyer payment must  be the same due to IC constraints. Therefore, this optimal payment must be Myerson's reserve price with respect to the value of that experiment. However, switching any experiment to a full information revelation experiment will never be worse. A formal argument is provided in  Appendix \ref{appendix:full-revealing}. 

Let $RevSingle^*$ denote the optimal revenue obtained in Proposition \ref{lem:full-revealing} without information discrimination, whereas $Rev^*$ denote the optimal revenue of Theorem \ref{thm:opt-scheme}. To understand  how much power information discrimination brings, we can study the ratio   $\frac{RevSingle ^*}{Rev^*} \in [0,1]$. Clearly, the larger this ratio is, the less crucial information discrimination is to revenue. 

It turns out that information discrimination is generally important for securing a high revenue. Specifically, in Appendix  \ref{appendix:bound-1}, we exhibit a concrete example showing that  the ratio   $\frac{RevSingle ^*}{Rev^*}$ can be arbitrarily close to $0$.  This is the case even when the value distribution of $e(t)$ is a regular distribution. 
 
Interestingly, it turns out that if the distribution of $e(t)$  defined in Equation \eqref{e(t)},  with randomness inherited from type $t$, has monotone hazard rate, then the optimal revenue without information discrimination can always guarantee at least a $1/e$ fraction of the optimal revenue. The proof of this proposition can be found in Appendix \ref{appendix:bound-2}.



\begin{proposition} \label{lem:bound-2}
If distribution of $e(t)$ has  monotone hazard rate (with randomness inherited from $t\sim F$),  then we always have $\frac{RevSingle ^*}{Rev^*} \geq \frac{1}{e}\, $. 
\end{proposition} 
\section{Proof of the Main Theorem }
In this section, we prove Theorem \ref{thm:opt-scheme}. Due to space limit, we will only provide a complete proof for Case 3.  The core idea for proving Case 1 and 2 is similar. We  thus defer them to Appendix \ref{sec:case1} and \ref{sec:case2}, respectively. 
The proof has two major steps: (1) characterizing useful properties of (any) feasible mechanisms; (2) leveraging the properties to derive the optimal mechanism. While the first step is also based on the analysis of the IC constraints as in   classic mechanism design, the conclusions we obtain are  quite different since our problem's constraints are different. Significantly deviating from the Myersonian approaches for classic mechanism design is our second main step, which arguably is much more involved due to additional constraints that we have to handle (this is also reflected in the more complex  format of our optimal mechanism).

\subsection{Useful Properties of Feasible Mechanisms}
\label{sec:proof:characterize}
Define \emph{feasible} mechanisms as the set of mechanisms $(\pi, p)$ that satisfy all the  constraints of program 
\eqref{lp:opt-formulation} (but not necessarily maximizing its objective). 
We first characterize the space of feasible mechanisms.    
To describe our characterization, it is useful to introduce the following quantity. 
\begin{equation} \label{P-def}
P_{\pi}(t) = \int_{q \in Q}  \pi(q, t)  g(q) \alpha(q)\,\dd q
\end{equation}
Note that $P_{\pi}(t)$ can be interpreted as the expected \emph{weighted} probability (with weight $\alpha(q)$) of being recommended the active action $1$. The following lemma summarizes our characterization. To illustrate the intuition, we only provide a proof of sufficiency here and defer the proof of necessity to Appendix \ref{appendix:feasible-M}.


\begin{lemma}[{\bfseries Characterization of Feasible Mechanisms}]
    A mechanism $(\pi, p)$ with non-negative payments is feasible if and only if it satisfies the following constraints:
     \begin{align}
     &   P_{\pi}(t) \text{ is monotone non-decreasing in } t \label{eq:signal-monotonicity} \\
     &  u(t) = u(t_1) +  \int_{t_1}^{t} P_{\pi}(x)\,\dd x, \forall t \in T \label{eq:buyer-utility-identify2} \\
     &  u(t_2) \geq v(t_2), \, \, \,  u(t_1) \geq 0 \label{eq:ir-t2} \\ 
     & p(t) \geq 0, \, \,  \forall t \in T \label{eq:non-negativity}
    \end{align}
\label{lem:feasible-M}
\end{lemma}


\begin{proof}[Proof of  Sufficiency]
We prove that constraints \eqref{eq:signal-monotonicity}--\eqref{eq:non-negativity} imply all the necessary constraints \eqref{cons:obedience},  \eqref{cons:IR} and \eqref{cons:IC}. The IC constraint \eqref{cons:IC} is equivalent to 
    \begin{equation*}
        u(t) \geq u(t') + \int_{q \in Q} \pi(q, t') \cdot g(q) [v(q,t) - v(q,t')]\,\dd q = u(t') + (t-t') P_{\pi}(t'). 
    \end{equation*}
    Therefore, constraints \eqref{eq:signal-monotonicity} and \eqref{eq:buyer-utility-identify2} imply the IC constraint \eqref{cons:IC} because if $t' < t$, we have 
    \begin{equation*}
        u(t) - u(t') = \int_{t'}^{t} P_{\pi}(x) \,\dd x \geq \int_{t'}^{t} P_{\pi}(t') \,\dd x = (t-t')P_{\pi}(t'). 
    \end{equation*}
    Similarly, when $t' > t$, we also have $    u(t) - u(t') \geq  (t-t')P_{\pi}(t')$. 
    
    The IR constraint \eqref{cons:IR} is equivalent to $u(t) \geq 0$ and $u(t) \geq v(t)$. Since $P_{\pi}(x)\geq 0$, constraint \eqref{eq:buyer-utility-identify2}, together with $u(t_1) \geq 0$, implies $u(t) \geq 0$ for any $t$. We now leverage $u(t_2) \geq v(t_2)$ to prove    $u(t) \geq v(t)$ for any $t$, as follows:
    \begin{gather*}
        u(t) =  u(t_1) + \int_{t_1}^{t} P_{\pi}(x)\,\dd x  
        =   u(t_2) - \int_{t}^{t_2} P_{\pi}(x)\,\dd x 
        \geq   v(t_2)  - \int_{t}^{t_2} P_{\pi}(x)\,\dd x.
    \end{gather*}
    Using the definition of $v(t_2)$ and $P_{\pi}(x)$, we get
    \begin{align*}
        u(t)=   &   \int_{q \in Q} g(q) \alpha(q)[t_2 + \beta(q)]\,\dd q  - \int_{t}^{t_2}   \int_{q \in Q}  \pi(q, x)  g(q) \alpha(q)\,\dd q \dd x  \\
        \geq  &   \int_{q \in Q} g(q) \alpha(q)[t_2 + \beta(q)]\,\dd q  - \int_{t}^{t_2}   \int_{q \in Q}   g(q) \alpha(q)\,\dd q \dd x  \\
        =  &   \int_{q \in Q} g(q) \alpha(q)[t + \beta(q)] \,\dd q  \\
        =& v(t) .
    \end{align*}
    
    Finally, the obedience constraint \eqref{cons:obedience} follows from the IR constraint \eqref{cons:IR} and $p(t) \geq 0$. 
\end{proof}

 Note that condition \eqref{eq:signal-monotonicity} is analogous to Myerson's allocation monotonicity condition in the auction design problem, but also differs in the sense that the value of an item in auction design only depends on the buyer type $t$ with no weight associated to it. In information selling, the value of taking the active action will  depend on the utility coefficient $\alpha(q)$.

Next we characterize the buyer's surplus $s(t) = u(t) - \max \{ 0, v(t) \}$, as expressed in Equation \eqref{eq:buyer-surplus},  from participating in the information selling mechanism.  
Recall that, with only the prior information,  a buyer of type $t$ has expected utility $v(t) = \int_{q \in Q} v(q, t) g(q)\,\dd q $ for the active action. Since $v(q, t)$ is monotone non-decreasing in $t$, we know that $v(t)$ is also monotone non-decreasing. Let $\bar{t}$ be any buyer  type at which $v(t)=0$. 
The following lemma characterize how the buyer's surplus changes as a function of his type. 
\begin{lemma}\label{lem:surplus-concave}
    Let $\bar{t}$ be any buyer type such that $v(\bar{t}) = \int_{q \in Q} v(q, \bar{t}) g(q) \,\dd q = 0 $. In any feasible mechanism $(\pi, p)$ with non-negative payments, the buyer's surplus $s(t) $ is 
    monotone non-decreasing for $t \in [t_1, \bar{t}]$ and monotone non-increasing for $t \in [\bar{t}, t_2]$.\footnote{$\bar{t}$ can be any one of them if there are multiple  $t$ such that $v(t)=0$. If no  $\bar{t}\in [t_1, t_2]$ makes $v(\bar{t})=0$, then either $v(t)<0$ or $v(t)>0$ for any $t\in T$ and in this case $s(t)$ is monotone within $T$. } 
\end{lemma}
\begin{proof}
    When $t \leq \bar{t}$, we have $v(t) \leq 0$. Therefore, without participating in the mechanism to purchase additional information, the buyer will get maximum utility $0$ by taking the passive action. So his surplus for participation is  $$s(t) = u(t) = u(t_1) + \int_{t_1}^{t} P_{\pi}(x)  \,\dd x$$ by the utility identify in Equation \eqref{eq:buyer-utility-identify2}. Since $u(t_1) \geq 0$ and $P_{\pi}(x) \geq 0$, it is easy to see that $s(t)$ is non-negative and monotone non-decreasing in $t$.  
    
    When $t \geq \bar{t}$, we have $v(t) \geq 0$. So the buyer's maximum utility is   $v(t)$ without participating in the information selling mechanism. We thus have 
    \begin{align*}
        s(t) =&  u(t) - v(t) \\
        =&  \left[ u(t_1) + \int_{t_1}^{t} \int_{q \in Q} \pi(q,x) \alpha(q) g(q)  \,\dd q \dd x  \right] - \left[ \int_{q \in Q}  \alpha(q) [t + \beta(q)] g(q) \,\dd q \right]  \\
        =&  \left[ u(t_1) + \int_{t_1}^{t} \int_{q \in Q} \pi(q,x) \alpha(q) g(q)  \,\dd q\dd x  \right] - \left[ \int_{t_1}^{t } \int_{q \in Q}  \alpha(q)    g(q) \,\dd q\dd x + v(t_1)  \right]  \\ 
        =&  u(t_1) - v(t_1) + \left[  \int_{t_1}^{t} \int_{q \in Q}  [\pi(q,x) - 1] \alpha(q)    g(q)  \,\dd q\dd x  \right]. 
    \end{align*} 
    Since $\pi(q,x) - 1 \leq 0 $ and $\alpha(q)    g(q) \geq 0$, we thus have that $s(t)$ is monotone non-increasing in $t$. Notably, $ s(t) \geq s(t_2) = u(t_2) - v(t_2) \geq 0$ by inequality \eqref{eq:ir-t2}. 
\end{proof}
\subsection{Deriving the Optimal Mechanism for Case 3}
\label{sec:case3}
With the characteristics of feasible mechanisms in  subsection \ref{sec:proof:characterize},  we are now ready to derive the optimal mechanism. This is where our proof starts to significantly deviate from standard approaches for classic mechanism design settings. To see the reasons, recall that Lemma \ref{lem:surplus-concave} shows that the buyer surplus  $s(t)$ in our problem generally increases first and then decreases. In    single-item auction design, however,  the buyer's utilities are always increasing in their types and thus   the optimal auction can always set the buyer's surplus to be $0$ at the lowest type \citep{myerson81}. In our case, however, both $s(t_1)$ and $s(t_2)$ could be the point with the lowest surplus, and we   have to figure out which one will be the lowest point under what conditions. Moreover,  the participation constraints require $u(t_2) \geq v(t_2)$ and $u(t_1) \geq 0$.\footnote{Generally,   the IR constraints require  $u(t) \geq \max \{ v(t), 0 \}, \forall t \in T$, but Lemma \ref{lem:feasible-M} reduces the IR constraints to $u(t_2) \geq v(t_2)$ and $u(t_1) \geq 0$.}  
To insure theses constraints, the format of the optimal mechanism and its derivation both become more involved.     

It turns out that whether the  minimum  buyer surplus will be achieved at point $t_1$ or point $t_2$ or simultaneously at both   $t_1$, $t_2$ depends on how large $v(t_1)$ and $v(t_2)$ are. 
Specifically, the optimal mechanism has different forms depending on whether $v(t_2)\le V_L$, $v(t_2)\ge V_H$, or $V_L< v(t_2)< V_H$, where $V_L$ and $V_H$ are defined in Equation \eqref{eq:upper-lower-bound1} and \eqref{eq:upper-lower-bound2}. 
To further illustrate these conditions,  the following lemma shows that the conditions for the above three cases  can   be equivalently expressed in terms of $v(t_1)$ as well. 

\begin{lemma}
    \label{lem:case_condition_t1}
    Define
    \begin{gather*}
        V'_L=-\int_{t_1}^{t_2} \int_{q: \beta(q) \geq - \underline{\phi}^+(x)} g(q) \alpha(q) \,\dd q\dd x\\
        V'_H=-\int_{t_1}^{t_2} \int_{q: \beta(q) \geq - \bar{\phi}^+(x)} g(q) \alpha(q) \,\dd q\dd x.
    \end{gather*}
    Then the three conditions $v(t_2)\le V_L$, $v(t_2)\ge V_H$, and $V_L< v(t_2)< V_H$ are equivalent to $v(t_1)\le V'_L$, $v(t_1)\ge V'_H$, and $V'_L< v(t_1)< V'_H$, respectively.
\end{lemma}
\begin{proof}
    We will only show that $v(t_2)\le V_L$ is equivalent to $v(t_1)\le V'_L$, as the other two cases follows from similar arguments.
    
    By definition, we have
    \begin{gather*}
        v(t_2)=\int_{ q \in Q}g(q)\alpha(q)[t_2+\beta(q)]\,\dd x=v(t_1)+(t_2-t_1)\int_{ q \in Q}g(q)\alpha(q)\,\dd q.
    \end{gather*}
    Thus $v(t_2)\le V_L$ can be written as:
    \begin{gather*}
        v(t_1)+(t_2-t_1)\int_{ q \in Q}g(q)\alpha(q)\,\dd q\le \max \{v(t_1), 0 \} + \int_{t_1}^{t_2} \int_{q: \beta(q) \geq -\underline{\phi}^+(x)} g(q) \alpha(q) \,\dd q\dd x.
    \end{gather*}
    Some re-arrangements yield:
    \begin{gather*}
        v(t_1)-\max \{v(t_1), 0 \}\le \int_{t_1}^{t_2} \int_{q: \beta(q) \geq -\underline{\phi}^+(x)} g(q) \alpha(q) \,\dd q\dd x-(t_2-t_1)\int_{ q \in Q}g(q)\alpha(q)\,\dd q,
    \end{gather*}
    which is equivalent to:
    \begin{gather*}
        \min\{v(t_1),0\} \le -  \int_{t_1}^{t_2} \int_{q: \beta(q) \leq -\underline{\phi}^+(x)}  g(q) \alpha(q) \,\dd q  \dd x=V'_L.
    \end{gather*}
    Note that the right-hand side is always non-positive. So the left-hand side has to be $v(t_1)$. Thus the condition $v(t_2)\le V_L$ is equivalent to $v(t_1)\le V'_L$, and also implies that $v(t_1)\le 0$.
\end{proof}

In the remainder of this section,  we will focus on the case with $V_L< v(t_2)< V_H$. For convenience of reference, we re-state the Case 3 of Theorem \ref{thm:opt-scheme} in the following proposition.   

\begin{proposition}
    \label{lem:case_3}
    If $V_L < v(t_2) < V_H$, let $c \in (0,1) $ be a constant that satisfies
    \begin{gather}
        \int_{t_1}^{t_2}\int_{q:\beta(q)\ge-\phi_c^+(t)} g(q)  \alpha(q)\,\dd q \dd t = v(t_2), \label{eq:choice_c}
    \end{gather}
    where $\phi_c^+(t)$ is the ironed version of the mixed virtual value function $\phi_c(t)$.
    Then the  threshold mechanism  with threshold signaling function  $\theta^*(t)   =-\phi_c^+(t)$ and the following payment function represents an optimal mechanism:
    \begin{gather*}
        p^*(t) = \int_{q\in Q}  \pi^*(q,t)g(q)  v(q,t) \,\dd q -  \int_{t_1}^{t} \int_{q \in Q}  \pi^*(q, x) g(q) \alpha(q) \,\dd q\dd x.
    \end{gather*}        
    Moreover,  $p^*(t)$ is non-decreasing in $t$ when $F(t)\leq c$ and monotone non-increasing when $F(t)>c$.   
\end{proposition}
Before proving the optimality of our mechanism, we   first argue that the constant $c$ described in Proposition \ref{lem:case_3} actually exists and thus the mechanism is well-defined. 
\begin{lemma}
    \label{lem:existence_of_C}
    If $V_L < v(t_2) < V_H$, there exists a constant $c\in (0,1)$ that satisfies Equation \eqref{eq:choice_c}.
\end{lemma}
\begin{proof}
    Lemma \ref{lem:case_condition_t1} implies that the condition $v(t_2)<V_H$ is equivalent to the following:
    \begin{gather}
        v(t_1) < -  \int_{t_1}^{t_2} \int_{q: \beta(q) \leq -\bar{\phi}^+(t)}  g(q) \alpha(q) \,\dd q  \dd t.
    \end{gather}
    The right-hand side of the above inequality is clearly non-positive. Thus $v(t_1)\le 0$ and $\max \{v(t_1), 0 \}=0$. The condition $V_L < v(t_2) < V_H$ can be written as:
    \begin{gather*}
        \int_{t_1}^{t_2} \int_{q: \beta(q) \geq -\underline{\phi}^+(t)} g(q) \alpha(q) \,\dd q\dd t < v(t_2) <  \int_{t_1}^{t_2} \int_{q: \beta(q) \geq -\bar{\phi}^+(t)} g(q) \alpha(q) \,\dd q\dd t.
    \end{gather*}
    When $c = 0$, we have $-\underline{\phi}^+(t)=-\phi_c^+(t)$ and
    \begin{gather}
        \int_{t_1}^{t_2}\int_{q:\beta(q)\ge-\phi_c^+(t)} g(q)  \alpha(q)\,\dd q \dd t = \int_{t_1}^{t_2} \int_{q: \beta(q) \geq -\underline{\phi}^+(t)} g(q) \alpha(q) \,\dd q\dd t < v(t_2).
        \label{eq:c=0}
    \end{gather}
    When $c = 1$, we have $-\bar{\phi}^+(t)=-\phi_c^+(t)$ and
    \begin{gather}
        \int_{t_1}^{t_2} \int_{q: \beta(q) \geq -\phi_c^+(t)} g(q) \alpha(q)\,\dd q\dd t = \int_{t_1}^{t_2} \int_{q: \beta(q) \geq -\bar{\phi}^+(t)} g(q) \alpha(q) \,\dd q\dd t > v(t_2).
        \label{eq:c=1}
    \end{gather}    
    Now we show that the following function is continuous in $c$
    \begin{gather*}
        \int_{t_1}^{t_2} \int_{q: \beta(q) \geq - \phi^+_c(t)} g(q) \alpha(q) \,\dd q\dd t.
    \end{gather*}
 





Specifically, we show that $\phi_c^+$ under $l_1$ norm $\int_{t_1}^{t_2} |\phi_c^+(t)| \dd t$ is continuous in $c$. Note that this is not obvious since the ironing procedure involves taking derivatives, which is not a continuous operator in general.\footnote{For example, the function sequence $\{ x^n \}_{n=1}^{\infty}$ tends to constant function $0$ but their derivatives do not. } Fortunately, continuity turns out to hold in our specific problem.   

Let $h_c(z)=\phi_c(F^{-1}(z))$ be the corresponding function defined in the ironing procedure (Appendix \ref{append:ironing-procedure}) where $z=F(t)$.  First, we observe that $h_c(z)$ is continuous in $c$ (all functions in this proof are under the $l_1$ norm),
because 
\begin{align}
\lim_{\epsilon \to 0}  \int_{0}^{1} | h_{c+\epsilon}(z) - h_c(z)| \dd z & =   \lim_{\epsilon \to 0}  \int_{t_1}^{t_2}  f(t) |\phi_{c+\epsilon}(t) - \phi_c(t)| \dd t\nonumber\\
&= \lim_{\epsilon \to 0} \int_{t_1}^{t_2} f(t) \left| t-\frac{c+\epsilon-F(t)}{f(t)} - t+\frac{c-F(t)}{f(t)}\right| \dd t  \nonumber  \\
&=\lim_{\epsilon \to 0}  \int_{t_1}^{t_2} |\epsilon| \dd t \nonumber \\
&= 0 .  \label{function-limit}
\end{align}
Next, we prove $l_c(t)$ is continuous in $c$.
By lemma \ref{lem:virtual_value_order},
 for any $0 \leq c < c' \leq 1$,  ${\phi}_c^+ (t) \geq {\phi}_{c'}^+(t)$ for any $t$. Thus, for any $0 \leq c < c' \leq 1$,  $l_c (z) \geq l_{c'}(z)$ for any $z$. 
Using this monotonicity and the fact that the ironing  procedure  satisfies
$  \int_0^1 h_c(z) \dd z =  \int_0^1 l_c(z) \dd z $, we have 
\begin{align*}
\lim_{\epsilon \to 0^-}  \int_{0}^{1} \left|l_{c+\epsilon}(z) - l_c(z)\right| \dd z & =   \lim_{\epsilon \to 0^-} \left[ \int_{0}^{1} l_{c+\epsilon}(z)  \dd z -   \int_{0}^{1} l_c(z) \dd z  \right] \\
& =\lim_{\epsilon \to 0^-} \left[  \int_{0}^{1} h_{c+\epsilon}(z)  \dd z  - \int_{0}^{1} h_c(z) \dd z \right]  \\
& =\lim_{\epsilon \to 0^-}  \int_{0}^{1} \left[h_{c+\epsilon}(z) - h_c(z) \right]\dd z \\
&= 0,
\end{align*}
where the last equation is due to the continuity of $h_c(z)$ in $c$ proved above. 
Similar derivation holds when $\epsilon \to 0^+$, so function $l_c(z)$  is continuous in $c$. 

Finally, it is straightforward to see that $\phi_c^+(t) = l_c(F(t))$ is continuous in $c$ as well because $f(t)$ has full support on $[t_1,t_2]$, and thus $  f_{\min}  \dd t \leq  \dd F(t)   \leq f_{\max} \dd t$ where $f_{\min}, f_{\max}$ are the smallest and largest value of the $f(t)$ on interval $[t_1,t_2]$. This concludes the argument that $\phi_c^+$ is continuous in $c$  under $l_1$ norm. Thus, we can conclude that the function $\int_{t_1}^{t_2} \int_{\beta(q) \geq - \phi^+_c(t)} g(q) \alpha(q) \,\dd q\dd t$ is continuous in $c$. Combined with Equation \eqref{eq:c=0} and \eqref{eq:c=1}, we must have $c\in(0,1)$ that satisfies Equation \eqref{eq:choice_c}.

\end{proof}

We   remark that the proof of Lemma \ref{lem:existence_of_C} relies on the assumption that the distribution of $\beta(q)$ does not contain a point mass. If this non-degeneracy assumption does not hold, we can slightly adjust our analysis to still obtain an optimal threshold mechanism but with randomized signals at only the boundary of the threshold experiments.  For completeness, we derive the optimal mechanism for this general case in Appendix \ref{appendix:partial_recommendation}. 

Lemma \ref{lem:existence_of_C} implies that the mechanism proposed in Proposition \ref{lem:case_3} exists. Next we show that it is also feasible.
\begin{lemma}
    \label{lem:case_3_feasible}
    The mechanism $({\pi}^*, {p}^*)$ defined according to $\phi_c^+(t)$ is feasible. Moreover, it satisfies: (1) $u(t_1) = 0, u(t_2) = v(t_2)$; (2)  $p^*(t)$ is non-decreasing in $t$ when $F(t)\leq c$ and monotone non-increasing when $F(t)>c$. 
\end{lemma}
\begin{proof}
    To prove Lemma \ref{lem:case_3_feasible}, it suffices to show that mechanism $({\pi}^*, {p}^*)$ satisfies all the constraints \eqref{eq:signal-monotonicity}, \eqref{eq:buyer-utility-identify2}, \eqref{eq:ir-t2}, and \eqref{eq:non-negativity} in Lemma \ref{lem:feasible-M}. By definition,
    \begin{gather*}
        P_{{\pi^*}}(t) = \int_{q:\beta(q)\ge -\phi_c^+(t)} g(q) \alpha(q) \,\dd q.
    \end{gather*}
    Since $\phi_c^+(t)$ is already ironed, it is non-increasing in $t$. Thus the integral domain of $P_{{\pi}^*}(t)$ gets larger as $t$ increases. So $P_{{\pi}^*}(t)$ is non-decreasing since $g(q)\alpha(q)\ge 0$ and thus satisfies constraint \eqref{eq:signal-monotonicity}.
    
    To show that the mechanism satisfies constraint \eqref{eq:buyer-utility-identify2}, note that the payment function in Proposition \ref{lem:case_3} implies 
    \begin{gather*}
        u(t)=\int_{q \in Q}   g(q) \pi^*(q,t)v(q,t_1) \,\dd q -p(t)=\int_{t_1}^{t} P_{{\pi}^*}(x)\,\dd x.
    \end{gather*}
We then have $u(t_1)=0$ and  consequently  $
        u(t)=u(t_1)+\int_{t_1}^{t} P_{{\pi}^*}(x)\,\dd x$ as in constraint \eqref{eq:buyer-utility-identify2}. 
    As for constraint \eqref{eq:ir-t2}, we already have $u(t_1)=0$. And
    \begin{gather*}
        u(t_2) = \int_{t_1}^{t_2} P_{{\pi}^*}(x)\,\dd x = \int_{t_1}^{t_2} \int_{q:\beta(q)\ge-\phi_c^+(t)} g(q) \alpha(q) \,\dd q \dd t = v(t_2),
    \end{gather*}
    where the last equality follows from the definition of the constant $c$.
    
More involved is to show the stated properties of the payment function and this is intrinsically related to the obedience constraints. We now argue that $p^*(t)$ is monotone non-decreasing when $F(t)\le c$, and monotone non-increasing when $F(t)\ge c$ (recall that $c\in (0,1)$).

 Let $t_c$ be the buyer type such that $F(t_c)  = c$. By Lemma \ref{lem:virtual_value_order}, we have  $\phi_c^+(t)\leq t$ when $ \forall F(t)\leq c$ and $\phi_c^+(t)\geq t$ when $F(t)\geq c$.   We first consider the case of $F(t) \leq c$, i.e., $t<t_c$. Let $t'$ be any number in the interval $[\phi_c^+(t), t]$. Thus $ \phi_c^+(t') \leq \phi_c^+(t)\le t$. And
    \begin{align*}
        &p^*(t)-p^*(t')\\
        =&\int_{q\in Q} g(q) \pi^*(q,t)v(q,t)\,\dd q -\int_{q\in Q} g(q) \pi^*(q,t')v(q,t')\,\dd q -  \int_{t'}^{t} P_{\pi^*}(x)  \,\dd x\\
        =&\int_{q:\beta(q)\ge -\phi_c^+(t)}g(q)v(q,t)\,\dd q-\int_{q:\beta(q)\ge -\phi_c^+(t')}g(q)v(q,t')\,\dd q-  \int_{t'}^{t} P_{\pi^*}(x)  \,\dd x.
    \end{align*}
    When $\beta(q)\ge -\phi_c^+(t)$, we have $v(q,t')=\alpha(q)[t'+\beta(q)]\ge \alpha(q)[t'-\phi_c^+(t)]\ge 0$, where the last inequality is due to the choice of $t'$. So the second term in the above equation satisfies:
    \begin{align*}
        &\int_{q:\beta(q)\ge -\phi_c^+(t')}g(q)v(q,t')\,\dd q\\
        =&\int_{q:\beta(q)\ge -\phi_c^+(t)}g(q)v(q,t')\,\dd q-\int_{q:-\phi_c^+(t)\le\beta(q) < -\phi_c^+(t')}g(q)v(q,t')\,\dd q\\
        \le&\int_{q:\beta(q)\ge -\phi_c^+(t)}g(q)v(q,t')\,\dd q.
    \end{align*}
    Thus,
    \begin{align*}
        &p^*(t)-p^*(t')\\
        \ge&\int_{q:\beta(q)\ge -\phi_c^+(t)}g(q)v(q,t)\,\dd q-\int_{q:\beta(q)\ge -\phi_c^+(t)}g(q)v(q,t')\,\dd q-  \int_{t'}^{t} P_{\pi^*}(x)  \,\dd x\\
        =&\int_{q:\beta(q)\ge -\phi_c^+(t)}g(q)\alpha(q)(t-t')\,\dd q-\int_{t'}^{t} P_{\pi^*}(x)  \,\dd x\\
        =&(t-t')P_{\pi^*}(t)-\int_{t'}^{t} P_{\pi^*}(x)  \,\dd x\\
        \ge &0,
    \end{align*}
    where the last inequality is due to the monotonicity of $P_{\pi^*}(t)$.
    
    Therefore, the payment function $p^*(t)$ is monotone non-decreasing in the interval $[\underline{\phi}^+(t),t]$. Since the set of intervals $\{[\underline{\phi}^+(t),t]\mid t\in [t_1,t_c]\}$ covers $[t_1,t_c]$, we conclude that $p^*(t)$ is monotone non-decreasing in $[t_1,t_c]$.
    
    Using similar analyses, we can show that $p^*(t)$ is monotone non-increasing in the interval $[t_c,t_2]$. Therefore, to prove that $p^*(t)\ge 0$ for all $t\in T$, it suffices to show that $p^*(t_1)\ge0$ and $p^*(t_2)\ge 0$. Indeed, we have
    \begin{align*}
        p^*(t_1) = \int_{q \in Q} \pi^*(q,t_1) g(q)  v(q,t_1)\, \dd q - u(t_1) = \int_{q:\beta(q)\geq-\phi_c^+(t_1)}  g(q)  v(q,t_1)\, \dd q \geq 0.
    \end{align*}
    The last inequality is because when $\beta(q)\ge-\phi_c^+(t_1)\ge -t_1$, we have $v(q,t_1)=\alpha(q)[t_1+\beta(q)]\ge 0$. And
    \begin{align*}
        p^*(t_2) =& \int_{q \in Q} \pi^*(q,t_2) g(q)  v(q,t_2)\, \dd q - u(t_2)\\
        =&\int_{q: \beta(q) \geq -\phi_c^+(t_2) } g(q) \alpha(q)[t_2+\beta(q)]\, \dd q -\int_{q \in Q} g(q) \alpha(q)[t_2 + \beta(q)]\,\dd q\\
        =&-\int_{q: \beta(q) < -\phi_c^+(t_2) } g(q) \alpha(q)[t_2+\beta(q)]\, \dd q\\
        \ge& 0,
    \end{align*}
    where the last inequality is because $\beta(q) < -\phi_c^+(t_2)\le -t_2$.
\end{proof}

Finally, we prove  the optimality of mechanism $(\pi^*,p^*)$. Since the optimal mechanism of  Proposition \ref{lem:case_3} depends on the ironed mixed virtual value functions $\phi_c^+(t)$ for all $c\in[0,1]$, our derivation here has to  employ the ironing trick for $\phi_c$ as well. We will first derive two equivalent representations of the revenue as a function of any feasible mechanism, and then interpolate  these two functions which give rise to the mixed virtual value. Finally,  we use the Myersonian approach  to argue that the defined mechanism $(\pi^*,p^*)$ in Proposition \ref{lem:case_3}  maximizes all terms in the revenue function simultaneously. 
\begin{proof}[Proof of Proposition \ref{lem:case_3}]
    Let $(\pi, p)$ be any feasible mechanism. We can write the revenue of  the seller as:
    \begin{gather*}
        REV (\pi, p)=\int_{t_1}^{t_2} f(t)p(t)\,\dd t=\int_{t_1}^{t_2} f(t)\left[\int_{q \in Q} g(q)  \pi(q,t)v(q,t) \,\dd q -u(t) \right]\,\dd t.
    \end{gather*}
    Applying Equation \eqref{def-u(t)} and \eqref{eq:buyer-utility-identify2}, we get
    \begin{align*}
        REV (\pi, p)=&\int_{t_1}^{t_2} f(t)\left[\int_{q \in Q} g(q)\pi(q,t)v(q,t)\,\dd q -\int_{t_1}^{t} P_{\pi}(x)\, \dd x -u(t_1) \right]\,\dd t\\
        =&\int_{t_1}^{t_2} f(t)\left[\int_{q \in Q} g(q) \pi(q,t)v(q,t)\,\dd q \right]\,\dd t-\int_{t_1}^{t_2} \int_{t_1}^{t}f(t) P_{\pi}(x)\,\dd x\dd t -u(t_1)\\
        =&\int_{t_1}^{t_2} f(t)\left[\int_{q \in Q} g(q) \pi(q,t)v(q,t)\,\dd q \right]\,\dd t-\int_{t_1}^{t_2} \int_{x}^{t_2}f(t) P_{\pi}(x)\,\dd t\dd x -u(t_1)\\
        =&\int_{t_1}^{t_2} f(t)\left[\int_{q \in Q} g(q)  \pi(q,t)v(q,t)\,\dd q \right]\,\dd t-\int_{t_1}^{t_2} [1-F(x)]P_{\pi}(x)\,\dd x-u(t_1),
    \end{align*}
    where the third equation comes from switching the order of integration. Thus
    \begin{align}
        &REV (\pi, p)\nonumber \\
        =&\int_{q \in Q} g(q)\left[\int_{t_1}^{t_2} f(t) \pi(q,t)v(q,t)\,\dd q \right]\,\dd t\\
        &-\int_{t_1}^{t_2} [1-F(t)]\int_{q \in Q} g(q) \pi(q,t)\alpha(q) \dd q\dd t-u(t_1) \nonumber \\
        =&\int_{q \in Q} g(q)\left[ \int_{t_1}^{t_2} f(t)\pi(q,t) \left(v(q,t)-\alpha(q)\frac{1-F(t)}{f(t)}\right)\,\dd t \right]\,\dd q -u(t_1) \nonumber \\
        =&\int_{q \in Q} g(q)\left[ \int_{t_1}^{t_2} f(t)\pi(q,t) \alpha(q) \left[\underline{\phi}(t) + \beta(q)\right]\,\dd t \right]\, \dd q -u(t_1). \label{eq:revenue-1}
    \end{align}

   The derived revenue function above uses $u(t_1)$ as the ``reference'' points. Similarly, using a variant of Equation \eqref{eq:buyer-utility-identify2} $u(t) = u(t_2) -  \int_{t}^{t_2} P_{\pi}(x)\,\dd x $, we  can derive an alternative form of the revenue with   $u(t_2)$ as the reference point: 
    \begin{gather}
        REV(\pi, p)=\int_{q \in Q} g(q)\left[ \int_{t_1}^{t_2} f(t)\pi(q,t)\alpha(q) \left[ \bar{\phi}(t) + \beta(q)\right]\,\dd t \right]\,\dd q -u(t_2). \label{eq:revenue-2}
    \end{gather}
    
    Note that Equation \eqref{eq:revenue-1} and \eqref{eq:revenue-2} are just different representations of the (same) revenue of any feasible mechanism $(\pi,p)$. Thus any convex combination of them also represents the same revenue. Using the constant $c$ given in Proposition \ref{lem:case_3} as the convex coefficient, we have
    \begin{align*}
        REV(\pi, p)=&c \left[\int_{q \in Q} g(q)\int_{t_1}^{t_2} f(t)\pi(q,t) \alpha(q) \left[\underline{\phi}(t) + \beta(q)\right]\,\dd t \, \dd q -u(t_1) \right] \\
        &+ (1-c)  \left[\int_{q \in Q} g(q) \int_{t_1}^{t_2} f(t)\pi(q,t)\alpha(q) \left[ \bar{\phi}(t) + \beta(q)\right]\,\mathrm{d}t  \mathrm{d}q -u(t_2) \right] \\
        = & \int_{t_1}^{t_2}\int_{q\in Q} \left[\phi_c(t) + \beta(q)\right] \pi(q,t) f(t) g(q) \alpha(q)\,\dd q\dd t - c u(t_1) - (1-c)u(t_2).
    \end{align*}
    
Next we employ the ironing trick. Define $h(z)=\phi_c(F^{-1}(z)), \forall z\in [0,1]$ where $F^{-1}(z)$ is the inverse function of CDF $F(t)$; let $ H(z)=\int_{0}^{z}h(r)\,\dd r$, $ L(z)$ be the convex hull of $H(z)$ and $l(z)=L'(z)$. We have $h_c(F(t))=\phi_c(t)$ and $l_c(F(t))=\phi_c^+(t)$ after ironing. So the first term in the right-hand side of the above equation can be written as
    \begin{align*}
        &\int_{t_1}^{t_2}\int_{q\in Q} \left[\phi_c(t) + \beta(q)\right] \pi(q,t) f(t) g(q) \alpha(q)\,\dd q\dd t\\
        =&\int_{t_1}^{t_2}\int_{q\in Q} \left[\phi_c^+(t) + \beta(q)\right] \pi(q,t) f(t) g(q) \alpha(q)\,\dd q\dd t \\
        &+\int_{t_1}^{t_2}\int_{q\in Q} \left[h_c(F(t)) - l_c(F(t))\right] \pi(q,t) f(t) g(q) \alpha(q)\,\dd q\dd t.
    \end{align*}
    Using integration by parts, we can simplify the second term as follows:
    \begin{align*}
        & \int_{t_1}^{t_2}\int_{q\in Q} \left[h_c(F(t)) - l_c(F(t))\right] \pi(q,t) f(t) g(q) \alpha(q)\,\dd q\dd t \\
        =& \int_{t_1}^{t_2} \left[h_c(F(t))- l_c(F(t))\right] P_{\pi}(t) \,\dd F(t) \\
        =& \left.\left[H_c(F(t))- L_c(F(t))\right] P_{\pi}(t) \right|_{t_1}^{t_2}  - \int_{t_1}^{t_2} \left[H_c(F(t))- L_c(F(t))\right] \,\dd P_{\pi}(t).
    \end{align*}
    Because $L_c$ is the ``convex hull'' of $H_c$, so $L_c(0) = H_c(0)$ and $L_c(1) = H_c(1)$. Thus the first term above is simply $0$. Therefore, we have 
    \begin{align}
        REV(\pi, p)
        = &\int_{t_1}^{t_2}\int_{q\in Q} \left[\phi_c^+(t) + \beta(q)\right] \pi(q,t) f(t) g(q) \alpha(q)\,\dd q\dd t \nonumber\\ 
        & - \int_{t_1}^{t_2} \left[H_c(F(t))- L_c(F(t))\right] \,\dd P_{\pi}(t)- c u(t_1) - (1-c)u(t_2).\label{eq:new-rev-3}
    \end{align}
    
  We argue that our feasible mechanism $({\pi}^*, {p}^*)$ simultaneously maximizes all the terms  in Equation \eqref{eq:new-rev-3}. Firstly, since ${\pi}^*(q, t)=1$ if and only if $\phi_c^+(t) + \beta(q)\ge 0$, $({\pi}^*, {p}^*)$ maximizes the first term. Secondly,  $({\pi}^*, {p}^*)$ also satisfies $u(t_1)=0$ and $u(t_2)=v(t_2)$   as shown in  Lemma \ref{lem:case_3_feasible}. Since $u(t_1)\ge 0$ and $u(t_2)\ge v(t_2)$ holds for any feasible mechanism as shown in Lemma \ref{lem:feasible-M}, $({\pi}^*, {p}^*)$ also maximizes the last two terms. Thirdly,  for the second term, note that $H_c(F(t))- L_c(F(t))\ge 0$ by definition, and $\dd P_{\pi}(t)\ge0$ for any feasible mechanism. Thus this term is always non-negative. However, we claim that with mechanism $({\pi}^*, {p}^*)$, this term is actually 0, i.e., the maximum possible. Clearly, the only interesting case is when $H_c(F(t))- L_c(F(t))> 0$. In this case $t$ must lie in an ironed interval $I$ and thus the convex hull $L_c(z)$ of $H_c(z)$ is linear in the ironing interval. This implies $l_c(z) = \phi_c^+(t)$ (where $z=F(t)$) is a constant and thus
    $
        P_{{\pi}^*}(t)=\int_{ q :\beta(q)\ge -\phi_c^+(t)}g(q)\alpha(q)\,\dd q
$
    is also constant in the interval $I$, leading to $\dd P_{{\pi}^*}(t) = 0$. 
    
    To summarize, the mechanism $({\pi}^*, {p}^*)$ optimizes all the 4 terms in Equation \eqref{eq:new-rev-3} simultaneously, thus is an optimal feasible mechanism.    
\end{proof}

\section{Generalizations}
\subsection{Generalized Utility Function}  \label{generalized-utility}
So far we have derived the optimal mechanism and its properties with  value functions that are linear and monotone non-decreasing in $t$, i.e., 
$    v(q,t)=\alpha(q)(t+\beta(q))$ for some $\alpha(q)\geq 0$. 
In this section we discuss how our analysis can be easily generalized to any value function that satisfy the following two assumptions:
\begin{assumption}[\textbf{Convexity and Monotonicity}]
For any $q$, $v(q,t)$ is convex and monotone non-decreasing in $t$.
\label{assum1:mono}
\end{assumption}
\begin{assumption}[\textbf{Monotone Virtual Values}]
For any $q\in Q $ and $ c\in [0,1]$, $\phi_c(t) = \frac{v(q,t)}{v'_t(q,t)} -\frac{c-F(t)}{f(t)} $ is non-decreasing in $t$ where   $v'_t(q,t)=\frac{\partial v(q,t)}{\partial t}$. 
\label{assum2:mono-ratio}
\end{assumption}

Our proof techniques can be applied in almost the same way with the above assumptions, and the threshold structure of the optimal mechanism will also remain similar. Assumption \ref{assum1:mono}  is the primary assumption; it retains  the monotonicity as  assumed before, but generalizes the linear value assumption to the much relaxed requirement of  convex values.  
Convexity is needed to preserve the equivalence between the monotonicity of $P_{\pi}(t)$ and IC constraints as in Lemma \ref{lem:feasible-M}, whereas the monotonicity of $v(q,t)$ in $t$  guarantees that the buyer's surplus will increase first and then decrease, and thus the participation constraint will bind only at type $t_1$ or $t_2$.  Assumption \eqref{assum2:mono-ratio} is a technical assumption which is only needed to avoid the ironing procedure so that the point-wise maximizing threshold mechanism still satisfies the monotonicity of $P_{\pi}(t)$ required by any feasible mechanism. We remark that under the widely adopted log-concavity assumption of type distribution $F(t)$,  we have $\frac{c-F(t)}{f(t)}  $ is non-increasing in $t$ for any $ c\in [0,1]$ \citep{Prkopa1971LogarithmicCM}. Therefore, to satisfy Assumption \ref{assum2:mono-ratio}, we only need an additional assumption that   the ratio $\frac{v(q,t)}{v'_t(q,t)}$ is  non-decreasing in $t$ for any $q$.

We make a few remarks about the generalized analysis. First, the threshold of the optimal mechanism will now depend on a natural generalization of the previous virtual value functions: $\frac{v(q,t)}{ v'_t(q,t)} -\frac{1-F(t)}{f(t)}$ or   $\frac{v(q,t)}{v'_t(q,t)} + \frac{F(t)}{f(t)}$ or their mixture. 
Second, it turns out that, with general value function, the four constraints listed in Lemma \ref{lem:feasible-M} are only necessary conditions but are \emph{no longer sufficient} for feasible mechanisms. However, it becomes a sufficient condition after we augment  these four conditions with   an additional requirement, i.e., the experiments are monotone in $t$ in the  sense that  $\forall q, \, \pi(q,t)$ is non-decreasing in $t$. To resolve this issue, we will relax the design space by considering $(\pi, p)$ that satisfies the four necessary constraints of Lemma \ref{lem:feasible-M}. This guarantees that any feasible mechanism is under our consideration since these constraints are necessary for feasibility, however we may suffer the risk of arriving at an infeasible mechanism. Noteworthily, the optimal solution of this relaxed optimization problem under Assumption \ref{assum1:mono} and \ref{assum2:mono-ratio} is a threshold mechanism which satisfies the monotone experiment requirement, i.e.,  $\forall q, \, \pi^*(q,t)$ is non-decreasing in $t$. This thus closes the gap between necessity and sufficiency, and shows that the mechanism we obtain is indeed a feasible mechanism. Since the detailed derivation for general value function is almost exactly the same as linear utility functions, up to the above two major differences, we omit them in this paper. 


\subsection{Correlated State and Buyer Type } \label{section-dependence}
Finally, we discuss how our results could be \emph{partially} generalized to the setting with correlated state $q$ and buyer type $t$.  This setting turns out to require more much careful treatment. First, within the general class of sequential mechanisms as we consider in this work, \citep{Babaioff12} show a similar result as \citep{cremer1988full} for auction design that the optimal mechanism can extract full surplus. However, the full-surplus-extracting optimal mechanism has to use  negative payments in order to guarantee that the payment from each buyer type is properly enforced even after they see the realized experiment outcomes.\footnote{Specifically, since the buyer is free to leave the mechanism after seeing an experiment outcome, the full-surplus-extracting   mechanism has to ask for a large upfront deposit at the beginning and then return the leftover of the deposit after deducting a buyer's payment for the realized experiment outcome.}  Notably, this is in contrast to the independent case, for which  our Lemma \ref{lem:positive-pay} shows that the optimal mechanism can always without loss of generality  use non-negative payments.  

Second, suppose negative payments are explicitly forbidden under correlated state and type, prior works gave examples showing that multiple rounds of information revelation can lead to strictly better revenue than any mechanism with a single round of information revelation, regardless whether the experiment outcomes can be contracted  \citep{Babaioff12} or cannot be contracted \citep{bergemann2018design}. However,  the design of optimal sequential mechanisms turns out to be quite challenging and, to our knowledge, is unknown in general. \cite{bergemann2018design}  restrict  their analysis to the design space of one-round mechanisms.  Towards this end, our Theorem \ref{thm:opt-scheme} can be generalized towards a characterization of the optimal mechanism for correlated $q,t$, but only within the space of \emph{one-round mechanisms} with positive payments.  
Specifically,  with correlated $q,t$, we will  need to instead impose   Assumption \ref{assum1:mono} and \ref{assum2:mono-ratio}  on the function $v(q,t)\mu(q,t)$, where $\mu(q,t)$ is the joint distribution of $q,t$, since they always bind together in all derivations.  All our derivations for Theorem \ref{thm:opt-scheme} can then be generalized in a straightforward way and is thus omitted here.   





\bibliographystyle{ecta}
\bibliography{ref} 

\clearpage
\section*{ONLINE APPENDIX}

\appendix

\section{Omitted Proofs in Section \ref{sec:model}}
\subsection{Proof of Lemma \ref{lem:revelation} (Revelation Principle)}\label{append:revelation}
\begin{proof}
Analogous to the proof of the classic revelation principle,  here we also construct a one-round mechanism and show that the constructed mechanism yields the same expected utility and revenue for both the buyer and the seller, respectively.


For any voluntary sequential mechanism, suppose that each seller node $n$ in the original mechanism is associated with distribution $\psi_n$ and each transfer node $n$ is associated with payment $p(n)$.
Let $\eta_n(c, t)$ be the buyer's optimal strategy in the above sequential mechanism, i.e., $\eta_n(c, t)$ is the probability of a buyer of type $t$ choosing child node $c$ at buyer node $n$. Let $L$ be the set of all leaf nodes, and $Z(l|q,t)$ is buyer type $t$'s belief about the probability of the game ending up at leaf node $l$ conditioned on the realized state $q$ if the buyer $t$ uses strategy $\eta_n(c, t)$. Denote by $\tau(l,t)$ the total payment made by a buyer of type $t$ along the way from the root of the game tree to leaf node $l$. We construct a one-round mechanism as follows:
\begin{enumerate}
    \item The seller asks the buyer to report his type $t$.
    \item The seller charges the buyer $p(t)=\Ex_{q\sim g}\left[ Z(l|q,t)\tau(l,t) \right]$
    \item Let $\Sigma = L$ and set $\pi(l|q,t)=Z(l|q,t)$. 
\end{enumerate}

Now consider the buyer's strategy in the above new mechanism, which now is only to report a type. First, we claim that if the buyer reports his type truthfully, his expected utility is exactly the same as that in the original mechanism. To see this, it suffices to show that the buyer: (1) obtains the same   information; (2) pays the same amount in expectation. Claim (1) holds because  the buyer updates his belief about the state $q$ each time he receives a message from the seller until he reaches a leaf $l$, and  the signal $l$ in the constructed new one-round mechanism will give him exactly the same posterior belief as the one he would get if he ends up in the leaf node $l$ in the original mechanism. Consequently, the buyer will derive the same value from revealed information. The later claim (2)  is by definition of the payment in the new mechanism. 

Second, and more importantly, we claim that if a buyer of  type $t$ misreports his type as $t'$ in the new mechanism, he also obtains the same utility as that of playing according to $t'$'s strategy $\eta_n(c,t')$ in the original mechanism. This is due to the following two observations.    Firstly, it is clear that he obtains the same information by construction of the new mechanism. Secondly, he also experiences the same expected payment  if misreporting $t'$. This crucially relies on the independence assumption. Specifically, since $t$ is independent of $q$,    the expected payment $p(t')=\Ex_{q\sim g}\left[ Z(l|q,t')\tau(l,t') \right]$ perceived by $t'$ himself is the same as the payment perceived by type $t$ when $t$ misreports $t'$, since they hold the same belief about $q$, which is always drawn from $g$.\footnote{This ceases to be true for correlated $q,t$, which is why Lemma \ref{lem:revelation} does not hold there. Specifically, when $t,q$ are correlated,  buyer type $t$ and $t'$ have different   beliefs about the probability of $q$. The payment for $t'$ now becomes  $\Ex_{q\sim g|t'}\left[ Z(l|q,t')\tau(l,t') \right]$ (where $q\sim g|t'$   is drawn from a conditional distribution after conditioning on $t'$), which is different from what buyer type $t$ thinks about his payment $\Ex_{q\sim g|t}\left[ Z(l|q,t')\tau(l,t') \right]$   when he misreports $t'$, since $q\sim g|t$ is now drawn from distribution conditioned on $t$.     
} 

Consequently, the new one-round mechanism is incentive compatible because if reporting $t'$ is more profitable for a buyer of type $t$ in the new mechanism,  so is it in the original mechanism. Then the strategy $\eta_n(c, t')$ would have been a better strategy for a buyer of type $t$ than $\eta_n(c, t)$, which    contradicts  the optimality of $\eta_n(c, t)$. Since the original mechanism is voluntary, so the new mechanism is IR. Finally, it is straightforward to see that the new one-round mechanism yields the same expected revenue, concluding the proof.  
\end{proof}

\subsection{Proof of Lemma \ref{lem:positive-pay} (Non-Negative Payment)}\label{appendix:positive-pay}
\begin{proof}
Let $(\pi, p)$ be any IC, IR  and one-round optimal mechanism. We construct a different mechanism $(\pi^*, p^*)$ which  satisfies the same constraints and remains optimal but with $p^*_t \geq 0$ for any $t$. For convenience, we divide buyer types into two sets: $T^+ = \{ t\in T: p_t \geq 0 \}$ is the set of types who have non-negative payments in mechanism  $(\pi, p)$ and $T^- = T \setminus T^+$ is the set of types who have negative payments. 

The mechanism $(\pi^*, p^*)$ is constructed from $(\pi, p)$ as follows: 
\begin{enumerate}
    \item  The mechanism for any $t \in T^+$ remains the same: for any $t\in T^+$,  let $p^*_t = p_t$ and $\pi^*_t = \pi_t$ for all $q \in Q$;
    \item The mechanism for any $t \in T^-$ becomes no information and no payment:  for any $t \in T^-$, let $p^*_t = 0$, and  $\pi^*_t$ be the mechanism that reveals no information (e.g., always sending a single signal).  
\end{enumerate} 
We observe that the constructed mechanism $(\pi^*, p^*)$ has three useful properties: (1) it yields revenue at least that of $(\pi, p)$ by construction; (2) all buyer types' payments are non-negative now; (3) individual rationality constraint is satisfied for every buyer type. The third property follows from the construction: the utility of any buyer type $t \in T^+$ did not change and the utility of a type $t \in T^-$ now pays $0$ and receives no information, so IR constraint is always  satisfied.   

However, the major issue with the constructed mechanism $(\pi^*, p^*)$   is that it may not be incentive compatible, i.e., bidder type $t$ may want to misreport $t'$. We first observe that the IC constraint for any $t \in T^+$ remains satisfied. First of all, any type $t \in T^+$ would not have incentive to deviate to another type $t' \in T^+$ due to the original IC constraint of $(\pi, p)$ and the fact that the mechanism for types in $T^+$ remains the same. 
We claim that any type $t \in T^+$ would not have incentive to deviate to a type $t'$ in $T^-$ as well. This is because compared to the original mechanism, the information obtained by mis-reporting $t' \in T^-$ is less (since \Seller{} reveals no information now) and the payment is more (since $p^*_{t'} = 0 > p_{t'}$). Therefore, if in mechanism $(\pi, p)$ buyer type $t$ does not have incentives to deviate to $t'$, he remains truthful in $(\pi^*, p^*)$.  

However, buyer type $t \in T^-$ may indeed have incentive to deviate to some type $t' \in T^+$ now, since they may want to receive beneficial information under some amount of payment.  
Here comes our last step of the construction --- adjusting  the above $(\pi^*, p^*)$  to make any type $t \in T^-$ to also satisfy IC without decreasing the revenue neither violating the IR and obedient constraint.  
To do so, for any $t \in T^-$, let $t' \in T^+$ be the most profitable deviation of type $t$, i.e., the deviation that maximizes type $t$'s utility. We adjust $(\pi^*, p^*)$ simply by adopting the scheme of type $t'$ to the type $t$ --- i.e.,  resetting $\pi^*_t = \pi_{t'}$ and $p^*_t = p_{t'}$. After such adjustment, the IC constraint for any type $t \in T^-$ is satisfied by construction because each of these types has indeed their most profitable mechanism. Meanwhile, this will also maintain the IC constraint for any type $t \in T^+$ since the adjustment did not add more entries to the menu. 
Note that IR constraint remains satisfied since the utility of any type $t \in T^+$  is non-decreasing in his adjustment. The revenue did not decrease as the payment $p^*(t)$ did not decrease in our adjustment for any $t \in T^+$. The only non-obvious part to verify is the obedience constraint. Indeed, the obedience constraint may be violated for type $t \in T^-$ during this adjustment since the  recommended optimal action for the $t' \in T^+$ might not be optimal for $t$. To achieve obedience, we simply ``rename'' the recommended action for $t$ to be the actual optimal action. This restores the obedience constraint for $t$. Note that, this will either not change the revealed information or lead to less revealed information (when type $t$'s optimal actions are the same under $\pi(\cdot, t')$), and thus will not hurt the IC constraints.  
\end{proof}

\section{ Ironing}
\label{append:ironing}

\subsection{Formal Description of the Ironing Procedure}\label{append:ironing-procedure}
\begin{definition}[\textbf{Ironing} \citep{myerson81}] \label{def:ironing}
    Let $t$ be the buyer's type with CDF $F(t)$ and PDF $f(t)$, and $\phi(t)$ be any function of the type $t$, called a \emph{virtual value function}. The \emph{ironed} function $\phi^+(t)$ can be obtained through the following process:
    \begin{enumerate}
        \item Let $z=F(t)$ be another random variable and define $h(z)=\phi(F^{-1}(z))$, where $F^{-1}(z)$ is the inverse function of $F(t)$.
        \item Define $H:[0,1]\mapsto \mathbb{R}$ to be the integral of $h(z)$:
            \begin{gather*}
                H(z)=\int_{0}^{z}h(r)\,\dd r.
            \end{gather*}
        \item Define $L:[0,1]\mapsto \mathbb{R}$ be the ``convex hull'' of function $H$:
            \begin{align*}
                L(z)=\min_{z_1,z_2,\gamma}\{ \gamma H(z_1)+(1-\gamma) H(z_2) \},
            \end{align*}
        where $z_1,z_2,\gamma\in[0,1]$ and $\gamma z_1+(1-\gamma)z_2=z$.
        \item Let $l(z)$ be the derivative of $L$:
            \begin{gather*}
                l(z)=L'(z).
            \end{gather*}
        \item Obtain $\phi^+(t)$ by variable substitution:
            \begin{gather*}
                \phi^+(t)=l(z)=l(F(t)).
            \end{gather*}
    \end{enumerate}
\end{definition}

The above ironing trick is widely used in the literature. 
Myerson's original work \citep{myerson81} only considers ironing for the lower virtual value function $\underline{\phi}(t) = t - \frac{1-F(t)}{f(t)}$. However, this procedure generalizes to any virtual value function (see also \citep{toikka2011ironing}).  





\subsection{Proof of   Lemma \ref{lem:virtual_value_order}: Useful Properties of Mixed Virtual Values}
\label{appendix:virtual_value_order_proof}

\subsubsection{Proof of the First Claim}
{\bf Restatement of the Claim.}  For any $0 \leq c < c' \leq 1$,  ${\phi}_c^+ (t) \geq {\phi}_{c'}^+(t)$ for any $t$. 
 
\begin{proof}
 We first prove the monotonicity of $\phi_c^+(t)$ in $c$. We will instead consider the function $\phi$ in a new variable space $z = F(t) \in [0,1]$, as opposed to the original space of $t$. Let $l_c(z) = \phi_c^+(F^{-1}(z))$ and $h_c(z) = \phi_c(F^{-1}(z))$. By definition of the ironing procedure, we have that $L_c(z) =  \int_{0}^z l_c(r) \, \dd r$   is the ``convex hull'' of $H_c(z) =  \int_{0}^z h_c(r) \, \dd r$.

Note that during the ironing of the smooth function $H_c(z)$ (since distribution $f(t)$ is assumed to be smooth), it divides the compact variable space $[0,1]$ into a countably many small intervals with breaking points $0=z_0, z_1, z_2, \cdots$. For any such interval $[z_i, z_{i+1}]$: either (1)  $H_c(z) = L_c(z)$ for any $z \in [z_i, z_{i+1}]$; or (2) $l_c(z)$ is a constant and $H_c(z) \geq L_c(z)$ for any $z \in [z_i, z_{i+1}]$. In the later case, we will call $[z_i, z_{i+1}]$ an \emph{ironing interval} and say $H_c(z)$ is at \emph{ironing state} in this interval.  We call $z_i$ the \emph{ironing starting point} and  $z_{i+1}$ the \emph{ironing ending point}. Similarly, in the former case, we call $[z_i, z_{i+1}]$ an \emph{non-ironing interval} and say $H_c(z)$ is at \emph{non-ironing state} in this interval. Note that in this case, $z_i$ will be an ironing ending point and $z_{i+1} $ will be an ironing starting point. In fact, in the sequence $0=z_0, z_1, z_2, \cdots$, ironing starting and ending points show up alternately.  The following are a few useful properties that will be needed.
\begin{enumerate}
    \item If $z \not = 0, 1$ is an ironing starting or ending point, then $h_c(z) = l_c(z)$ and $H_c(z) = L_c(z)$. 
    \item if $[z_i, z_{i+1}]$ is an ironing interval for $H_c(z)$, then we have $l_c(z) = \frac{H_c(z_{i+1}) - H_c(z_i)}{z_{i+1} - z_i}$ for any  $z \in [z_i, z_{i+1}]$ 
    \item For any $z \in [0,1]$  we have  $h_c(z) = t+\frac{F(t)}{f(t)}-\frac{c}{f(t)}  \geq  t+\frac{F(t)}{f(t)}-\frac{c'}{f(t)} = h_{c'}(z)$  where $t = F^{-1}(z)$.
    \item Due to Property (3) above,  for any $z$  we have  $H_c(\bar{z}) - H_c(z) \geq    H_{c'}(\bar{z}) - H_{c'}(z)$ for any $\bar{z} > z$. Moreover,  $H_c(0)  =    H_{c'}(0) = 0$. 
\end{enumerate}

Similarly, we can also have a sequence of ironing starting and ending points for the function $H_{c'}(z)$. Let us merge all the ironing starting and ending points of $H_c(z)$ and $H_{c'}(z)$ together, and re-order them as $0 = z_0, z_1, \cdots$. Notably, within any interval $[z_i, z_{i+1}]$, both function $H_c(z)$ and $H_{c'}(z)$ can only have a single state, either the ironing state or the non-ironing state. 

We first prove $l_c(0) \geq l_{c'}(0)$. This follows from a case analysis about whether $0$ is an ironing ending or starting point for $H_c$.\footnote{That ``$0$ is an ironing ending point'' effectively means $0$ is not ironed. This is to make the argument to be in one-to-one correspondence with later arguments for points inside the type interval. }
\begin{itemize}
    \item If $0$ is an ironing ending point for both $H_c, H_{c'}$, meaning both functions are not in the ironing state at $0$ and its neighborhood, we know $l_c(0) = h_c(0) \geq h_{c'}(0) =   l_{c'}(0)$, as desired.  
    \item If $0$ is an ironing ending point only for  $H_c$ but an ironing starting point for $H_{c'}$, this means  $H_c$  is  in the non-ironing state at $0$ and its neighborhood whereas $H_{c'}$ is in the ironing state. Then we have $l_c(0) = h_c(0) \geq h_{c'}(0)   \geq l_{c'}(0)$, as desired. 
    \item If $0$ is an ironing starting point for $H_c$ (it does not matter it is an ironing ending or starting point for $ H_{c'}$), this means $H_c$     is  on ironing state at $0$ and its neighborhood. Let $\bar{z} \geq z_1$ be the immediate next ironing ending point for $H_c$. Suppose, for the sake of contradiction, that $l_c(0) < l_{c'}(0)$. We thus have 
    \begin{align*}
    H_c(\bar{z}) - H_c(0) &= l_c(0) \cdot (\bar{z} - 0)  & \mbox{by Property (2) above } \\
    & <  l_{c'}(0) \cdot (\bar{z} - 0)  & \mbox{by assumption} \\ 
    & \leq  L_{c'}(\bar{z}) - L_{c'}(0)   & \mbox{by convexity of $L_{c'}$}  \\
     & \leq  H_{c'}(\bar{z}) - H_{c'}(0)   & \mbox{$L_{c'}(0) = H_{c'}(0), L_{c'}(\bar{z}) \leq  H_{c'}(\bar{z})$}  
    \end{align*} 
     This contradicts Property (4) above.  
    Therefore, we must  $l_c(0)   \geq    l_{c'}(0)$, as desired.  
\end{itemize}

Next, we will prove that for any $i=0,1,\cdots$ and any interval   $[z_i, z_{i+1}]$ ---  more conveniently denoted as $[a,b]$ with interval lower bound $a$ and upper bound $b$ ---  we will have $l_c(z) \geq l_{c'}(z)$ for any $z \in [a,b]$. 

Our proof uses an induction argument over the intervals indexed by $i$. Specifically, suppose we already have $l_c(a) \geq l_{c'}(a)$, we will show $l_c(z) \geq l_{c'}(z)$ for any $z \in [a,b]$. This, together with the base case for $a=0$ as proved above, will prove the monotonicity of $l_c(z)$ on $c$.  

The proof  uses a case analysis about whether the ending point $b$ of the interval $[a,b]$ is an ironing starting point or ending point for $H_c$ or for $H_{c'}$. Note that there will be four cases here. This is because we do not know whether $b$ is an ironing point for $H_c$ or $H_{c'}$ and thus have to consider both possibilities. Here, we will use the crucial property that both  $H_c$ and $H_{c'}$ will have the same state, i.e., ironing or non-ironing state, within $[a,b]$ due to our choice of $a,b$. 
\begin{itemize}
    \item If  $b$ is an ironing ending point for function $H_{c'}$, we have for any $z \in [a,b]$ 
      \begin{align*}
     l_c(z) & \geq l_c(a)  & \mbox{by convexity of $L_c$ } \\
    & \geq   l_{c'}(a)  & \mbox{by induction hypothesis} \\ 
    &  =  l_{c'}( z)    & \mbox{$H_{c'}$ is at ironing state in $[a,b]$}
    \end{align*} 
    
    \item If  $b$ is an ironing starting point for function $H_{c'}$,  this means  $H_{c'}$ is in the non-ironing state within $[a,b]$. If $H_c$ is also in the non-ironing state within $[a,b]$, then we have $l_c(z) = h_c(z) \geq h_{c'}(z) = l_{c'}(z)$ as desired. Now we consider the case that $H_c$ is in the ironing state within $[a,b]$. Let $\bar{z} \geq b$ be the immediate next ironing ending point for $H_c$. Suppose, for the sake of contradiction, that $l_c(z) < l_{c'}(z)$ for some $z\in [a,b]$. Since $H_c$ is in the ironing state within $[a,\bar{z}]$, we know that $l_c(r) = l_c(z)  < l_{c'}(z) \leq l_{c'}(r)$ for any $r \in [b, \bar{z}]$ since $l_{c'}(z)$ is monotone non-decreasing in $z$. We have  
    \begin{align*}
    H_c(\bar{z}) - H_c(b) & \leq  L_c(\bar{z}) - L_c(b) & \mbox{$L_{c}(\bar{z}) = H_{c}(\bar{z}), L_{c}(b) \leq  H_{c}(b)$} \\ 
    & \leq l_c(b) \cdot (\bar{z} - b)  & \mbox{$H_c$ is in the ironing state in $[a,\bar{z}]$} \\
    & <  l_{c'}(b) \cdot (\bar{z} - b)  & \mbox{by assumption} \\ 
    & \leq  L_{c'}(\bar{z}) - L_{c'}(b)   & \mbox{by convexity of $L_{c'}$}  \\
     & \leq  H_{c'}(\bar{z}) - H_{c'}(b)   & \mbox{$b$ is an ironing starting point for  $H_{c'}$}  
    \end{align*} 
      This contradicts Property (4) above.  
    Therefore, we must have $l_c(z)   \geq    l_{c'}(z)$ for all $z \in [a,b]$, as desired. Note that one corner case for this situation is when $\bar{z}$ happens to equal $b$, i.e., $b$ is both the ironing starting point of $H_{c'}$ and ironing ending point of $H_c$. Our argument above does not apply to this corner situation since the strict ``$<$'' above becomes ``$=$''. However, this corner case can  be proved via a simpler argument: $\forall z\in[a,b]$, $l_c(z) = l_c(b) = h_c(b) \geq h_{c'}(b) \geq h_{c'}(z) = l_{c'}(z)$ where the second equality is due to the fact that $b$ is an ironing ending point of $H_c$ and the last equality is due to the fact that  $H_{c'}$ is in the non-ironing state within $[a,b]$. 
    
    \item If  $b$ is an ironing starting point for function $H_{c}$,  this means  $H_{c}$ is in the non-ironing state within $[a,b]$. If $H_{c'}$ is also in the non-ironing state within $[a,b]$, then we have $l_c(z) = h_c(z) \geq h_{c'}(z) = l_{c'}(z)$ as desired. If $H_{c'}$ is   in the ironing state within $[a,b]$, then we have $l_c(z)   \geq l_c(a) \geq \ l_{c'}(a) = l_{c'}(z)$ as desired. 
    
    \item Finally, if   $b$ is an ironing ending point for function $H_{c}$,  this means  $H_{c}$ is in the  ironing state within $[a,b]$. If $H_{c'}$ is also in the ironing state within $[a,b]$, then we have  $l_c(z)   = l_c(a) \geq l_{c'}(a) = l_{c'}(z)$ as desired. If $H_{c'}$ is  in the non-ironing state within $[a,b]$, then we have $l_c(z) = l_c(b) = h_c(b) \geq h_{c'}(b) \geq  l_{c'}(z)$ where: (1) the first equality is because $H_c$ is in the ironing state within $[a,b]$; (2) the second equality is because $b$ is an ironing ending point for $H_c$ and (3) the last inequality is because $H_{c'}$ is in the non-ironing state within $[a,b]$ and thus $h_{c'}(z) = l_{c'}(z)$ is monotone non-decreasing in $z$.  
\end{itemize}

\end{proof}

\subsubsection{Proof of the Second Claim}
{\bf Restatement of the Claim. }  For any $c \in [0 ,1]$, let  $t_c$ be the buyer type such that $F(t_c)  = c$. Then we have $\phi_c^+(t)\leq t, \forall t\leq t_c$ and $\phi_c^+(t)\geq t, \forall t\geq t_c$. This also implies $ \underline{\phi}^{+}(t)< t< \bar{\phi}^{+}(t) , \forall t \in (t_1, t_2)$. 
 
\begin{proof}

Let $H_c$ and $L_c$ be the corresponding functions defined in Definition \ref{def:ironing} when ironing the mixed virtual value function ${\phi}_c(t)$. Let $I=(a,b)\subset [0,1]$ be any ironed interval, which thus satisfies $H_c(a)=L_c(a)$ and $H_c(b)=L_c(b)$ but $H_c(z)>L_c(z)$ for all $z\in I $. Since     the type $t$'s distribution $f(t)$ is smooth, the interval $[0,1]$ can be partitioned into countably many sets of disjoint ironed and non-ironed intervals.  

Let  $t_c$ be the unique buyer type such that $F(t_c)  = c$ (uniqueness by non-negativity of the density function). Crucial to this proof is to argue that $c$ \emph{cannot} be in any ironed interval. Suppose, for the sake of contradiction, that $c \in (a,b)$ belongs to an ironed interval $(a,b)$. Let $t_a,t_b$ be such that $F(t_a) = a$ and $F(t_b)= b$.  By definition of ironing, function $L_c$ is linear in $(a,b)$ and their derivatives are constants satisfying $\phi_c^{+}(t_a)= \phi_c^{+}(t_c)= \phi_c^{+}(t_b)$. However, $\phi_c(t_a) = t_a - \frac{c-F(t_a)}{f(t_a)} = t_a - \frac{c-F(t_a)}{f(t_a)} \leq t_a $ whereas  $\phi_c(t_b)  = t_b - \frac{c-F(t_b)}{f(t_b)} \geq t_b $. Since $a,b$ are the boundary of the ironed interval $(a,b)$, we have $ \phi^+_c(t_a) = \phi_c(t_a) \leq t_a < t_b \leq \phi_c(t_b) = \phi^+_c(t_b)$, which contradicts $\phi^+_c(t_a)  = \phi^+_c(t_b)$. This shows that $c$ cannot be within any ironed interval. 


Now we consider any type $t \in [t_1, t_2]$.  If its corresponding $z=F(t)$ falls into a non-ironed interval, then we have $H_c(z^*)=L_c(z^*)$ for all $z^*$ in the same interval. So $\phi_c^{+}(t)=L_c'(z)=H_c'(z)=\phi_c(t)$. Consequently, when $t \leq t_c$, we have $\phi_c(t) = \phi_c^{+}(t) = t - \frac{c-F(t)}{f(t)} \leq t$ whereas  when $t \geq t_c$, we have $\phi_c(t) = \phi_c^{+}(t) = t - \frac{c-F(t)}{f(t)} \geq t$, as desired.  

If  $z=F(t)$ falls into an ironed interval $I=(a,b)$. Since     $c$ cannot be within any ironed interval, $I$ can either be fully on the left-hand side of $c$ or fully on the right-hand side. 
\begin{enumerate}
\item When the boundaries satisfy $a\leq b \leq c$,  for any $t$ with $F(t) \in (a,b)$, we have  $\phi_c^{+}(t) = \phi_c^{+}(t_a)  = \phi_c(t_a) = t_a - \frac{c - F(t_a)}{f(t_a)} \leq t_a \leq t$. 
\item When the boundaries satisfy $a\geq b \geq c$, for any $t$ with $F(t) \in (a,b)$, we have $\phi_c^{+}(t) = \phi_c^{+}(t_b)  = \phi_c(t_b) =t_b - \frac{c - F(t_b)}{f(t_b)} \geq t_b \geq t$. 
\end{enumerate}

Finally,  by plugging $c=1$ and $c=0$, we get ${\phi}_1^+ (t) = \underline{\phi}^{+}(t)< t< \bar{\phi}^{+}(t) ={\phi}_0^+ (t) , \forall t \in (t_1, t_2)$.

\end{proof}

 \section{ Revenue without Price Discrimination}
\subsection{Proof of Proposition \ref{lem:full-revealing}} \label{appendix:full-revealing}
 \begin{proof}
 Let $\pi^*, p^*$ be any   incentive compatible optimal mechanism with a single experiment. IC implies that $p^*(t)$ must be the same for all types since otherwise all buyer types would report the same type $t' = \argmax_{t}p^*(t)$ to get the minimum payment as they all get the same information from $\pi^*$ anyway. Therefore, the optimal mechanism boils down to a pricing mechanism with price $p^*(t)$ for experiment $\pi^*$. 
 
 Now consider the full information experiment  denoted by  $\bar{\pi}$. By Blackwell's order of information structure \citep{Blackwell53}, full information is more informative than any other signaling scheme and thus leads to higher buyer utility from decisions. Consequently, the mechanism $\bar{\pi}, p^*$ must obtain at least the revenue of the optimal mechanism $\pi^*, p^*$ since any buyer who is willing to buy under $\pi^*, p^*$ must also be willing to buy under  $\bar{\pi}, p^*$. Finally, the mechanism defined in the proposition obtains at least the revenue of $\bar{\pi}, p^*$ and thus must also be optimal. 
\end{proof}

\subsection{Arbitrarily Worse Revenue without Information Discrimination --- an Example } \label{appendix:bound-1}
 
Consider the instance which has  $v(q,t) = t-q$, $q \in [0,C]$ where $C > 8$, and $\forall q, \,g(q) = \frac{1}{C}$. With $t \in [\sqrt{2C},\frac{C}{2}]$, the CDF $F(t)$ is defined below and has a point-mass probability at $t = \frac{C}{2}$:
\begin{gather*}
F(t) =  
\begin{cases}
1 -\frac{2C}{t^2} & \text{if } t\in [\sqrt{2C},\frac{C}{2}) \\ 
1 & \text{if } t = \frac{C}{2} \\ 
\end{cases}
\end{gather*}
Thus, the PDF $f(t)$ of $t$ when $t \in [\sqrt{2C},\frac{C}{2})$ is $f(t)= \frac{4C}{t^3}$.
The low virtual value can be easily computed and is regular:
\begin{gather*}
\underline{\phi}(t)  =  
\begin{cases}
\frac{t}{2} & \text{if } t\in [\sqrt{2C},\frac{C}{2}) \\ 
t & \text{if } t = \frac{C}{2} \\ 
\end{cases}
\end{gather*}

With straight calculations, we can get 
$v(t_2) = 0 \leq V_L$.
Thus, this instance falls into case 1. Based on Theorem \ref{thm:opt-scheme}, 
the optimal revenue can be bounded by:
\begin{align*}
Rev^* &= 
\int_{t} f(t) p(t) dt
\geq \frac{1}{4} \ln \frac{C}{8} .
\end{align*}


Now we show the optimal revenue restricted to a single-entry menu $Rev^*_{single}$ for this instance. From Lemma \ref{lem:full-revealing}, we know the optimal revelation mechanism will be full revelation. Thus, we can view the single-entry menu problem as an simple posted price problem with buyer's utility function $e(t)$.
By calculation, we get
\begin{align*}
e(t) 
= \frac{t^2}{2C}, \quad
e'(t) &
= \frac{t}{C} 
\end{align*}

Since $C > 8$ and $t \in [\sqrt{2C},\frac{C}{2}]$,
$e(t)$ is increasing  in $t$.
Denote the CDF and the PDF of $v = e(t)$ by $H(v)$ and $h(v)$.
By monotonicity of $v=e(t)$ and $t\in [\sqrt{2C},\frac{C}{2})$, we have 
\begin{align*}
H(v) = F(t) = F(e^{-1}(v))  = 
\begin{cases}
1 - \frac{1}{v}  & \text{if } v\in [1,\frac{C}{8}) \\ 
1 & \text{if } v = \frac{C}{8} \\ 
\end{cases}
\end{align*}


Notably, $H(v)$ is an equal revenue distribution. Thus, the optimal revenue in the single menu setting  
\begin{gather*}
Rev^*_{single} = \max\left\{1,  \max_v \int_{v} [1-H(v)] v \, \dd v \right\}= 1    
\end{gather*}
is a constant for this instance. 
Noteworthily, $H(v)$ is regular but not MHR.


We now reach the conclusion, if the distribution $H(v)$ is not a MHR distribution, 
\begin{align*}
\lim_{C \rightarrow +\infty}  \frac{Rev_{single}^*}{Rev^*}  \leq   \lim_{C \rightarrow +\infty} \frac{1}{  \frac{1}{4} \ln \frac{C}{8}   } = 0.
\end{align*}

\subsection{Proof of Proposition \ref{lem:bound-2}} \label{appendix:bound-2}
\begin{proof}
The problem of selling information with a single experiment can be viewed as a single item auction problem with $e(t)$ as the bidder's type and $H(v)$ as the type distribution. From this perspective, $Rev^*_{single}$ is the optimal revenue with a single experiment obtained by Myerson's optimal reserve price.  Using $Wel^*$ as the optimal social welfare,
  \cite{10.1145/1807342.1807364}  show  that the optimal social welfare in any single item auction with one bidder is at most $e$ times of the optimal revenue. This implies $Wel^* \leq e Rev^*_{single}$. Since $Wel^*$ is the surplus of information buyer when revealing full information, $Rev^* \leq Wel^*$ because of the IR constraint. Thus, we reach the conclusion that
  \begin{align*}
Rev^* \leq Wel^* \leq e Rev^*_{single}.
 \end{align*}


\end{proof}

\section{Characterization of Feasible Mechanisms --- Proof of Lemma \ref{lem:feasible-M}}\label{appendix:feasible-M} 
In this appendix section, we show that the conditions in Lemma \ref{lem:feasible-M} are also necessary for any feasible mechanism. We start by analyzing the IC Constraints. First,  Constraint \eqref{cons:IC} can be re-arranged as follows:
\begin{equation*}
    \int_{q \in Q} [\pi(q, t) - \pi(q, t')] \cdot g(q) v(q,t) \,\dd q\geq p(t)  - p(t').
\end{equation*}
Therefore, the IC constraint implies the following two inequalities about any two types $t, t'$: 
\begin{gather}
    \int_{q \in Q} [\pi(q, t) - \pi(q, t')] \cdot g(q) v(q,t) \,\dd q \geq  p(t)  - p(t'), \label{eq:ic1}\\
    \int_{q \in Q} [\pi(q, t') - \pi(q, t)] \cdot g(q) v(q,t')\,\dd q \geq  p(t')  - p(t).\label{eq:ic2}
\end{gather}

Combining Inequality \eqref{eq:ic1} and \eqref{eq:ic2}, we obtain the following constraint for any pair of types $t, t'$:
\begin{align*}
    &\int_{q \in Q} [\pi(q, t') - \pi(q, t)] \cdot g(q) v(q,t) \,\dd q \\
    \leq &p(t')  - p(t) \\
    \leq &\int_{q \in Q} [\pi(q, t') - \pi(q, t)] \cdot g(q) v(q,t') \,\dd q . 
\end{align*}

Therefore, the right-hand side of the above inequality must be at least its left-hand side. This implies the following \emph{necessary condition} for any IC information selling mechanism $(\pi, p)$. That is, for any $t, t' \in T$, we have 
\begin{eqnarray}\nonumber
    0 &\leq&  \int_{q \in Q} [\pi(q, t') - \pi(q, t)] \cdot g(q) [v(q,t') - v(q, t)]\,\dd q \\  \label{eq:IC-outcome0}
    & = & [ t' - t] \int_{q \in Q} [\pi(q, t') - \pi(q, t)] \cdot g(q) \alpha(q)\,\dd q   .
\end{eqnarray}
Recall the definition of $ P_{\pi}(t)$ \eqref{P-def}
$$ P_{\pi}(t) = \int_{q \in Q}  \pi(q, t) \cdot g(q) \alpha(q)\,\dd q. $$
Note that $P_{\pi}(t)$ can be interpreted as the expected \emph{weighted} probability of being recommended the active action where the weights are $\alpha(q)$.  A simple case analysis  for $t' > t$ and $t' < t$ implies that   Inequality \eqref{eq:IC-outcome0} is equivalent to  that $P_{\pi}(t)$ is monotone non-decreasing in $t$.  
We thus term this the \emph{signaling monotonicity}. This is analogous to Myerson's allocation monotonicity condition as in auction design, but is different. Specifically, in Myerson's optimal auction, the value of an item directly depends on buyer type $t$ with no weight associated to it. In information selling, the value of taking the active action will  depend on the utility coefficient $\alpha(q)$. 

We now derive a relation between experiment $\pi$ and payment rule $p$ for any IC mechanism. We start by analyzing \Buyer's utility. Note that any buyer of type $t$ will derive a non-zero value only from the active action recommendation since  the passive action always leads to buyer value $0$. Therefore, as defined in \eqref{def-u(t)}, \Buyer{} of type $t$ has the following utility:  
\begin{gather*}
    \text{Utility of Buyer Type }t: \quad u(t)= \int_{q \in Q} \left[ g(q) \pi(q,t)v(q,t) \right] \,\dd q -p(t)
\end{gather*}

Re-arranging Inequality \eqref{eq:ic1}, we have 
\begin{align*}
    u(t) =& \int_{q \in Q} \left[ g(q) \pi(q,t)v(q,t) \right] \,\dd q -p(t)  \\
    \geq& \int_{q \in Q} \left[ g(q) \pi(q,t')v(q,t) \right] \,\dd q -p(t')\\  
    =& \int_{q \in Q} \left[ g(q) \pi(q,t')v(q,t) \right] \,\dd q + u(t') - \int_{q \in Q} \left[ g(q) \pi(q,t') v(q,t') \right] \,\dd q \\ 
    =& \int_{q \in Q} \left[ g(q) \pi(q,t')[ v(q,t) - v(q, t')]  \right] \,\dd q + u(t')  \\
    =& (t-t') P_{\pi}(t') + u(t'). 
\end{align*}

As a result, Inequality \eqref{eq:ic1} implies  $u(t) -u(t') \geq (t-t')P_{\pi}(t') $. Together with a similar derivation from Inequality \eqref{eq:ic2}, we  have the following inequality 
\begin{gather*}
    (t-t') P_{\pi}(t')  \le u(t)-u(t')\le (t-t') P_{\pi}(t).
\end{gather*}
Note that the above inequality holds for any $t, t'$. Therefore, by letting $t'  \to t$ and invoking that fact that $P(t)$ is monotone and continuous, we can integrate the above equation from $t_1$ to $t$ and obtain  the inequalities: $$ \int_{t_1}^{t} P_{\pi}(x)  \,\dd x \leq u(t) - u(t_1) \leq   \int_{t_1}^{t} P_{\pi}(x) \,\dd x. $$ This implies the following:   
\begin{gather*} 
    u(t) =  u(t_1) +   \int_{t_1}^{t} P_{\pi}(x) \,\dd x.  
\end{gather*}
Note that both the signaling monotonicity and the above equation are the necessary outcomes of the incentive compatibility constraints, more precisely, the outcome of Constraints \eqref{eq:ic1} and \eqref{eq:ic2}.

\section{Optimal Mechanism for Case 1 ($v(t_2)\le V_L$)}\label{sec:case1}
In this section, we derive the optimal mechanism for the first case of Theorem \ref{thm:opt-scheme}. Similar to Section \ref{sec:case3}, we will first prove that our mechanism is feasible. Then we show it achieves the optimal revenue among all feasible mechanisms.

\begin{lemma}
    \label{lem:case1_feasible}
    The threshold mechanism $(\pi^*,p^*)$ defined according to $\underline{\phi}^+(t)$ is feasible.
\end{lemma}
\begin{proof}
    Using the characterization of Lemma \eqref{lem:feasible-M}, it suffices to show that the given mechanism satisfies Constraints \eqref{eq:signal-monotonicity}-\eqref{eq:non-negativity}. Since the ironed lower virtual value function $\underline{\phi}^+(t)$ is monotone non-decreasing, we know that the threshold $\theta^*_t =  -\underline{\phi}^+(t)$   is monotone non-increasing in $t$. This implies that 
    \begin{gather*}
        P_{{\pi}^*}(t) = \int_{q \in Q} {\pi}^*(q, t) g(q)\alpha(q)\,\dd q =   \int_{q: \beta(q) \geq -\underline{\phi}^+(t)}  g(q)\alpha(q) \,\dd q 
    \end{gather*}
    is monotone non-decreasing in $t$ since a larger $t$ leads to a smaller integral lower bound, satisfying Constraint \eqref{eq:signal-monotonicity}.
    
    The utility function is, by definition,
    \begin{gather*}
        u(t)  = \int_{q\in Q} [g(q) {\pi}^*(q,t)v(q,t)] \,\dd q - {p}^*(t)=\int_{t_1}^{t} P_{{\pi}^*}(x) \,\dd x
    \end{gather*}
    which implies $u(t_1)=0$, and
    \begin{gather*}
        u(t)  = u(t_1)+\int_{t_1}^{t} P_{{\pi}^*}(x) \,\dd x,
    \end{gather*}
    satisfying Constraint \eqref{eq:buyer-utility-identify2}.
    
    For Constraint \eqref{eq:ir-t2}, we already have $u(t_1)=0$. Now we prove $u(t_2) \geq v(t_2)$. Lemma \ref{lem:case_condition_t1} shows that the condition $v(t_2)\le V_L$ is equivalent to $v(t_1)\le V'_L$. Also, it is easy to see that $V'_L\le 0$, which implies $v(t_1)\le 0$. So $\max\{v(t_1), 0\}=0$, and
    \begin{align*}
        u(t_2) &= \int_{t_1}^{t_2}P_{{\pi}^*}(x) \,\dd x   = \int_{t_1}^{t_2} \int_{q: \beta(q) \geq -\underline{\phi}^+(t)} g(q) \alpha(q)\,\dd q \dd x \\
        &=  \int_{t_1}^{t_2} \int_{q: \beta(q) \geq -\underline{\phi}^+(t)} g(q) \alpha(q) \,\dd q \dd x + \max\left\{0,v(t_1) \right\} \\
        &\geq v(t_2).
    \end{align*}
    
    Finally, we argue that the payment is non-negative, i.e., Constraint \eqref{eq:non-negativity} is satisfied. By lemma \ref{lem:virtual_value_order}, we have for all $t\in T$, $\underline{\phi}^{+}(t)\leq t$.

    Let $t'$ be any number in the interval $[\underline{\phi}^+(t), t]$. Thus 
    %
    \begin{align*}
        &p^*(t)-p^*(t')\\
        =&\int_{q\in Q} [g(q) \pi^*(q,t)v(q,t)]\,\dd q -\int_{q\in Q} [ \pi^*(q,t')g(q)v(q,t')]\,\dd q -  \int_{t'}^{t} P_{\pi^*}(x)  \,\dd x\\
        =&\int_{q:\beta(q)\ge -\underline{\phi}^+(t)}g(q)v(q,t)\,\dd q-\int_{q:\beta(q)\ge -\underline{\phi}^+(t')}g(q)v(q,t')\,\dd q-  \int_{t'}^{t} P_{\pi^*}(x)  \,\dd x.
    \end{align*}
    When $\beta(q)\ge -\underline{\phi}^+(t)$, we have $v(q,t')=\alpha(q)[t'+\beta(q)]\ge \alpha(q)[t'-\underline{\phi}^+(t)]\ge0$, where the last inequality is because of the choice of $t'$. So the second term in the above equation satisfies:
    \begin{align*}
        &\int_{q:\beta(q)\ge -\underline{\phi}^+(t')}g(q)v(q,t')\,\dd q \\
        =&\int_{q:\beta(q)\ge -\underline{\phi}^+(t)}g(q)v(q,t')\,\dd q-\int_{q:-\underline{\phi}^+(t)\le\beta(q) < -\underline{\phi}^+(t')}g(q)v(q,t')\,\dd q\\
        \le&\int_{q:\beta(q)\ge -\underline{\phi}^+(t)}g(q)v(q,t')\,\dd q.
    \end{align*}
    Thus,
    \begin{align*}
        &p^*(t)-p^*(t')  \\
        \ge&\int_{q:\beta(q)\ge -\underline{\phi}^+(t)}g(q)v(q,t)\,\dd q-\int_{q:\beta(q)\ge -\underline{\phi}^+(t)}g(q)v(q,t')\,\dd q-  \int_{t'}^{t} P_{\pi^*}(x)  \,\dd x\\
        =&\int_{q:\beta(q)\ge -\underline{\phi}^+(t)}g(q)\alpha(q)(t-t')\,\dd q-\int_{t'}^{t} P_{\pi^*}(x)  \,\dd x\\
        =&(t-t')P_{\pi^*}(t)-\int_{t'}^{t} P_{\pi^*}(x)  \,\dd x\\
        \ge & 0,
    \end{align*}
    where the last inequality is due to the monotonicity of $P_{\pi^*}(t)$.

    Therefore, the payment function $p^*(t)$ is monotone non-decreasing in the interval $[\underline{\phi}(t),t]$. Since the set of intervals $\{[\underline{\phi}(t),t]\mid t\in T\}$ covers the interval $T$, we conclude that $p(t)$ is monotone non-decreasing in $T$. Therefore, to prove that $p(t)\ge 0$ for all $t\in T$, it suffices to show that $p(t_1)\ge0$. Indeed, we have
    \begin{align*}
        p^*(t_1) = \int_{q \in Q} \pi^*(q,t_1) g(q)  v(q,t_1)\, \mathrm{d}q - u(t_1) = \int_{q:\beta(q)\geq-\underline{\phi}^+(t_1)}  g(q)  v(q,t_1)\, \mathrm{d}q \geq 0.
    \end{align*}
    The inequality holds because when $\beta(q)\ge-\underline{\phi}^+(t_1)\ge -t_1$, we get $v(q,t_1)=\alpha(q)(t_1+\beta(q))\ge 0$.
\end{proof}

Now we prove that the mechanism defined according to $\underline{\phi}^+(t)$ is optimal, i.e., achieves the maximum possible revenue among all feasible mechanisms.

\begin{lemma}
    \label{lem:optimal-case1} 
    If $v(t_2) \leq V_L $, the threshold mechanism with  threshold signaling function $\theta^*(t) =  -\underline{\phi}^+(t) $ and the following payment function represents an optimal mechanism:
    \begin{gather*}
        p^*(t) = \int_{q\in Q}  \pi^*(q,t)g(q)  v(q,t) \,\dd q -  \int_{t_1}^{t} \int_{q \in Q}  \pi^*(q, x) g(q) \alpha(q) \,\dd q\,\dd x.
    \end{gather*}
    where $\pi^*$ is determined by $\theta^*(t)$ as in Definition \ref{def:threshold}. 
\end{lemma}
\begin{proof}
    
    
    According to the proof of Proposition \ref{lem:case_3}, the revenue of any feasible mechanism can be written as:
    \begin{align*}
        REV (\pi, p)=\int_{q \in Q} g(q)\left[ \int_{t_1}^{t_2} f(t)\pi(q,t) \alpha(q) \left[\underline{\phi}(t) + \beta(q)\right]\dd t \right]\dd q -u(t_1),
    \end{align*}
Let $\underline{H}(\cdot)$, $\underline{h}(\cdot)$, $\underline{L}(\cdot)$, and $\underline{l}(\cdot)$ the corresponding functions when ironing the virtual value $\underline{\phi}(t)$.
We can write the first term of the revenue function as follows:
\begin{align*}
    &\int_{q \in Q} \int_{t_1}^{t_2} \left[\underline{\phi}(t) + \beta(q)\right] f(t){\pi}(q,t)  g(q) \alpha(q) \,\dd t \dd q  \\
    = &\int_{q \in Q} \int_{t_1}^{t_2} \left[\underline{\phi}^+(t) + \beta(q)\right] f(t){\pi}(q,t)  g(q) \alpha(q) \,\dd t \dd q \\ 
    &+ \int_{q \in Q} \int_{t_1}^{t_2} [\underline{h}(F(t))- \underline{l}(F(t))] f(t){\pi}(q,t)  g(q) \alpha(q) \,\dd t \dd q.
\end{align*}
This is because by definition, $\underline{\phi}^+(t)=\underline{l}(F(t))$ and $\underline{\phi}(t)=\underline{h}(F(t))$. Using integration by parts, we can simplify the second term
\begin{align*}
    & \int_{q \in Q} \int_{t_1}^{t_2} [\underline{h}(F(t))- \underline{l}(F(t))] f(t){\pi}(q,t)  g(q) \alpha(q) \,\dd t \dd q \\
    =& \int_{t_1}^{t_2} [\underline{h}(F(t))- \underline{l}(F(t))] P_{{\pi}}(t) \,\dd F(t) \\
    =& [\underline{H}(F(t))- \underline{L}(F(t))] P_{{\pi}}(t) |_{t_1}^{t_2}  - \int_{t_1}^{t_2} [\underline{H}(F(t))- \underline{L}(F(t))] \,\dd P_{{\pi}}(t).
\end{align*}
Because $\underline{L}$ is the ``convex hull'' of $\underline{H}$ on $[0,1]$, $\underline{L}(0) = \underline{H}(0)$ and $\underline{L}(1) = \underline{H}(1)$. Thus the  term $ [\underline{H}(F(t))- \underline{L}(F(t))] P_{{\pi}}(t)|_{t_1}^{t_2}$ is simply $0$, and we have 
\begin{align*}
    REV ({\pi}, {p}) 
    = &\int_{q \in Q} \int_{t_1}^{t_2} \left[\underline{\phi}^+(t) + \beta(q)\right] f(t){\pi}(q,t)  g(q) \alpha(q) \,\dd t \dd q \nonumber \\ 
    & - \int_{t_1}^{t_2} [\underline{H}(F(t))- \underline{L}(F(t))] \,\dd  P_{{\pi}}(t) -u(t_1) .
\end{align*}

Now consider mechanism $({\pi}^*, {p}^*)$. $\pi^*$ maximizes the first term since ${\pi}^*(q,t) = 1, \forall q,t$ with $\beta(q)+\underline{\phi}^+(t)\ge 0$. Also, by definition, we have
\begin{gather*}
    u(t)=\int_{q\in Q} {\pi}^*(q,t)g(q) v(q,t) \,\dd q-p(t)=\int_{t_1}^{t}P_{{\pi^*}}(x) \,\dd x.
\end{gather*}
Thus we have $u(t_1)=0$. 

As for the second term, note that $\underline{H}(F(t))- \underline{L}(F(t))\ge 0$ by definition, and $\dd P_{\pi}(t)\ge0$ for any feasible mechanism. Thus the second term is always non-negative. However, we claim that with mechanism $({\pi}^*, {p}^*)$, this term is actually 0. The only interesting case is when $\underline{H}(F(t)) - \underline{L}(F(t)) > 0$. To prove our claim, it suffices to show that $\dd P_{{\pi}^*}(t) = 0$. In this case, $t$ must lie in an ironed interval $I$. Thus $\underline{L}(z)$ is linear in the interval $I$, where $z=F(t)$. This implies that $\underline{\phi}^+(t)=\underline{l}(z)=\underline{L}'(z)$ is constant. So
\begin{gather*}
    P_{{\pi}^*}(t)=\int_{ q \in Q}{\pi}^*(q,t)g(q)\alpha(q)\,\dd q=\int_{ q :\beta(q)\ge -\underline{\phi}^+(t)}g(q)\alpha(q)\,\dd q
\end{gather*}
is also constant in the interval $I$, which leads to $\dd P_{{\pi}^*}(t)$ being 0.

Therefore, mechanism $({\pi}^*, {p}^*)$ optimizes all terms in Equation \eqref{eq:revenue-1} simultaneously, hence optimal.
\end{proof}

Note that the above derivation of $REV(\pi,p)$ uses the equation $u(t) = \int_{t_1}^{t} P_{\pi}(x)\,\dd x+u(t_1)$ to expand $u(t)$ with $t_1$ as the reference point. This is also the original Myerson's approach. This approach works in Myerson's optimal auction design because there \Buyer{}'s surplus equals \Buyer{}'s utility from participating in the mechanism since the only outside option is to not purchase, resulting in utility $0$. Therefore, in Myerson's optimal auction design,  $u(t_1) \geq 0$   guarantees the IR constraint, i.e., $u(t) \geq 0$, for any feasible mechanism. This, however, ceases to be true in our setup because $s(t_1) \geq 0$ does not guarantee $s(t_2) \geq 0$. In fact, Lemma \ref{lem:surplus-concave} shows that $s(t)$ attains its maximum value at $\bar{t}$, where $\bar{t}$ is a zero of $v(t)$ function. Nevertheless, we know that the optimal mechanism must satisfy either $s(t_1) = 0$ or $s(t_2) = 0 $, since otherwise, we can shift the entire $s(t)$ curve down by a constant --- achieved by asking each buyer type to pay the same additional amount --- until one of them reaches $0$.

\section{Optimal Mechanism for Case 2 ($v(t_2)\ge V_H$)}
\label{sec:case2}
In this section, we will discuss the second case of our main result, i.e., when $v(t_2)\ge V_H$. In this case, if we still use $t_1$ as the reference point and follow the same analysis of Case 1, we will end up having a mechanism with $u(t_2)<v(t_2)$, hence infeasible. To solve this problem, we write the revenue expression $REV(\pi,p)$ using $t_2$ as the reference point. Although the resulting mechanism looks different, the approach for deriving it is quite similar to that in the proof of Case 1.

We still start with showing the feasibility of the given mechanism $(\pi^*,p^*)$.

\begin{lemma}
    The threshold mechanism $(\pi^*,p^*)$ defined according to $\bar{\phi}^+(t)$ is feasible.
\end{lemma}
\begin{proof}
    According to Lemma \eqref{lem:feasible-M}, it suffices to show that that the given mechanism satisfies Constraints \eqref{eq:signal-monotonicity}-\eqref{eq:non-negativity}. Since the ironed upper virtual value function $\bar{\phi}^+(t)$ is monotone non-decreasing, we know that the threshold ${\theta}^*(t) = -\bar{\phi}^+(t)$   is monotone non-increasing in $t$. This implies that 
    \begin{equation*}
        P_{\pi^*}(t) = \int_{q \in Q} \pi^*(q, t) g(q)\alpha(q) \,\dd q =   \int_{q: \beta(q) \geq -\bar{\phi}^+(t)}   g(q)\alpha(q) \,\dd q 
    \end{equation*}
    is monotone non-decreasing in $t$ since a larger $t$ leads to a larger $-\bar{\phi}^+(t)$ and thus larger integral domain for $q$. So Constraint \eqref{eq:signal-monotonicity} is satisfied.
    
    We now prove that $(\pi^*, p^*)$ satisfies Constraint \eqref{eq:buyer-utility-identify2}. Plugging the payment function $\pi^*(t)$ into the definition of $u(t)$, we get
    \begin{gather*}
        u(t)  = \int_{q\in Q} g(q) \pi^*(q,t)v(q,t) \,\dd q - p^*(t) =    v(t_2) - \int_{t}^{t_2} P_{\pi^*}(x) \,\dd x.
    \end{gather*}
    It is easy to see that $u(t_2) =  v(t_2)$, which can be plugged back to the above equality to obtain Constraint \eqref{eq:buyer-utility-identify2}.
    
    For Constraint \eqref{eq:ir-t2}, we already have $u(t_2) =  v(t_2)$. And
    \begin{align*}
        u(t_1) = v(t_2) - \int_{t_1}^{t_2} P_{\pi^*}(x)\,\dd x \geq \max \{v(t_1), 0 \} + \int_{t_1}^{t_2} P_{\pi^*}(x)\,\dd x - \int_{t_1}^{t_2} P_{\pi^*}(x) \,\dd x \geq 0,
    \end{align*}
    where the first inequality is due to the condition $v(t_2)\ge V_H$.
    
    Finally, we show that the payment $p^*(t)$ is non-negative i.e., $p^*(t)$ satisfies Constraint \eqref{eq:non-negativity}. By lemma \ref{lem:virtual_value_order}, we have for all $t\in T$, $ t \leq \bar{\phi}^{+}(t)$.

    For any $t>t_1$ and $t'\in [t, \bar{\phi}^+(t)]$, we have
    \begin{align}
        &p^*(t') -p^*(t) \nonumber \\
        &= \int_{q\in Q}  \pi^*(q,t')g(q)v(q,t') \,\dd q - \int_{q\in Q} g(q) \pi^*(q,t)v(q,t) \,\dd q - \int_{t}^{t'} P_{\pi^*}(x)  dx \nonumber\\ 
        &= \int_{q: \beta(q) \geq -\bar{\phi}^+(t')}  g(q) v(q,t') \,\dd q - \int_{q: \beta(q) \geq -\bar{\phi}^+(t)}  g(q) v(q,t) \,\dd q - \int_{t}^{t'} P_{\pi^*}(x)\,\dd x. \label{eq:payment_decreasing}
    \end{align}
    Observe that $\bar{\phi}^+(t)\leq \bar{\phi}^+(t')$ since $t\le t'$.
    So the first term in the right-hand side can be written as:
    \begin{align*}
        & \int_{q: \beta(q) \geq -\bar{\phi}^+(t')}  g(q) v(q,t') \,\dd q\\
        &= \int_{q: \beta(q) \geq -\bar{\phi}^+(t)}  g(q) v(q,t') \,\dd q+\int_{q: -\bar{\phi}^+(t')\le \beta(q) < -\bar{\phi}^+(t)}  g(q) v(q,t') \,\dd q.
    \end{align*}
    When $\beta(q) < -\bar{\phi}^+(t)$, we have $v(q, t')=\alpha(q)[t'+\beta(q)]\le \alpha(q)[t'-\bar{\phi}^+(t)]\le 0$, where the inequality is due to the choice of $t'$. Therefore, the second term in the right-hand side of the above equation is negative. As a result,
    \begin{gather*}
        \int_{q: \beta(q) \geq -\bar{\phi}^+(t')}  g(q) v(q,t') \,\dd q\le \int_{q: \beta(q) \geq -\bar{\phi}^+(t)}  g(q) v(q,t') \,\dd q.
    \end{gather*}
    Combined with Equation \eqref{eq:payment_decreasing}, we get
    \begin{align*}
        &p^*(t') -p^*(t) \\
        \le & \int_{q: \beta(q) \geq -\bar{\phi}^+(t)}  g(q) v(q,t') \,\dd q - \int_{q: \beta(q) \geq -\bar{\phi}^+(t)}  g(q) v(q,t) \,\dd q - \int_{t}^{t'} P_{\pi^*}(x)\,\dd x\\
        =&\int_{q: \beta(q) \geq -\bar{\phi}^+(t)}g(q)\alpha(q)(t'-t)\,\dd q- \int_{t}^{t'} P_{\pi^*}(x)\,\dd x\\
        =&(t'-t)P_{\pi^*}(t)-\int_{t}^{t'} P_{\pi^*}(x)\,\dd x\\
        \le&0.
    \end{align*}
    
    This shows that $p^*(t)$ is monotone non-increasing in the interval $[t,\bar{\phi}^+(t)]$ for any $t>t_1$. Since set of intervals $\{[t,\bar{\phi}^+(t)]\mid t\in T\}$ covers interval $T$, we can conclude that $p^*(t)$ is monotone non-increasing in the entire interval $T$.\footnote{Similar techniques are also used to proved existence of solutions for differential equations.} Thus, to show that the payment is always non-negative, we only need to prove that $p^*(t_2)\ge 0$. Indeed,
    \begin{align*}
        p^*(t_2) &= \int_{q \in Q} g(q) \pi^*(q,t_2)v(q,t_2)\,\dd q - v(t_2) + \int_{t_2}^{t_2} P_{\pi^*}(x)\,\dd x   \\
        &= \int_{q \in Q} g(q) \pi^*(q,t_2)  v(q,t_2)\, \dd q - \int_{q \in Q} g(q)  v(q,t_2)\, \dd q\\ 
        &=-\int_{q :\beta(q)<-\bar{\phi}^+(t_2)} g(q)  v(q,t_2)\, \dd q.
    \end{align*}
    When $\beta(q)<-\bar{\phi}^+(t_2)<-t_2$, we have $v(q,t_2)=\alpha(q)[t_2+\beta(q)]<0$. Thus $p^*(t_2)\ge 0$.
\end{proof}
\section{Threshold Mechanisms with Random Signals} 
\label{appendix:partial_recommendation}
We assumed that the probability distribution of $\beta$ does not have point masses in the main body of the paper. This is to ensure the existence of the constant $c$ in Case 3 of Theorem \ref{thm:opt-scheme}. But if the distribution of $\beta$ has point masses, such a $c$ may not exist. In this case, we will need to slightly modify our mechanism and incorporate random signals. If such a $c$ does not exist, it must be that both the distributions of $\beta$ and $\phi_c^+$ contains point masses, more specifically,  the measure of $\{(\beta, t)\mid \beta(q)=\phi_c^+(t)=\zeta \}$ is non-zero for some $\zeta$.




For any $c\in [0,1]$, let  $
    \phi_c(t)=c\underline{\phi}(t) + (1-c) \bar{\phi}(t)
$ be the mixed virtual value function 
and 
\begin{gather*}
    Y(c)=\int_{t_1}^{t_2} \int_{q: \beta(q) \geq -\phi_c^+(t)} g(q) \alpha(q) \,\dd q\dd t
\end{gather*}
be a function of $c$. 

We first prove the monotonicity of $Y(c)$.
As shown in Lemma \ref{lem:virtual_value_order},
$\forall 0 \leq c < c' \leq 1$, we have ${\phi}_c^+ (t) \geq {\phi}_{c'}^+(t)$ $\forall  t$. Thus, when $c$ is increasing, $\phi_c^+(t)$ is (weakly) decreasing and the threshold $-\phi_c^+(t)$ is   (weakly) increasing. So the function $Y(c)$ will integrate a non-negative function $g(q)\alpha(q)$ over a smaller region of $q$  and is (weakly) decreasing.

Next, we argue that $Y(c)$ is left-continuous. By monotoniciy, we know that $Y(c)$ is continuous almost everywhere. For any  $  c \in (0,1)$ and any arbitraryly small positive  $\epsilon$, we have
\begin{align*}
&\lim_{\beta \to c^-} Y(\beta) - Y(c)  \\
&=\lim_{\epsilon \to 0^+} Y(c-\epsilon) - Y(c) \\
&= \lim_{\epsilon \to 0^+} \int_{t_1}^{t_2} \int_{q: \beta(q) \geq -\phi_{c-\epsilon}^+(t)} g(q) \alpha(q) \,\dd q\dd t - \int_{t_1}^{t_2} \int_{q: \beta(q) \geq -\phi_c^+(t)} g(q) \alpha(q) \,\dd q\dd t \\
&= \lim_{\epsilon \to 0^+} \int_{t_1}^{t_2} \int_{q: -\phi_{c-\epsilon}^+(t) \leq \beta(q) < -\phi_{c}^+(t)} g(q) \alpha(q) \,\dd q\dd t \\
&= 0,
\end{align*}
where the last equation is because whenever there is a point mass such that  the measure of $\{(\beta, t)\mid \beta(q)=\phi_{\beta}^+(t)=\zeta \}$ is non-zero for some $  \beta < c$, we can always increase the lower bound of the integral to exclude this point mass by choosing an $\epsilon$ smaller than $c - \beta$. Consequently, we have  $
\lim_{\beta \to c^-} Y(\beta) = Y(c),
$
so function $Y(c)$ is left continuous on $(0,1)$.




%

Now, we are ready to define our signaling function for the case with point mass $\beta(q)$. Since $Y(c)$ is monotone (weakly) decreasing and is   left continuous, the following min is well-defined
\begin{align} \label{c-infimum}
c&=\max\{\beta \mid Y(\beta) \geq  v(t_2) \},   
\end{align}
and moreover we can use binary search to find the  $c$.

Given the above $c$, we define the following experiment. Define the following constant $D$, 
\begin{align*}
D &= \frac{v(t_2)-(Y(c) - \int_{t_1}^{t_2} \int_{q:    \beta(q)=-\phi_c^+(t) } g(q) \alpha(q) \,\dd q\dd t)}{\int_{t_1}^{t_2} \int_{q:    \beta(q)=-\phi_c^+(t) } g(q) \alpha(q) \,\dd q\dd t} \\
&= \frac{v(t_2)-\int_{t_1}^{t_2} \int_{q:    \beta(q)>-\phi_c^+(t) } g(q) \alpha(q) \,\dd q\dd t}{\int_{t_1}^{t_2} \int_{q:    \beta(q)=-\phi_c^+(t) } g(q) \alpha(q) \,\dd q\dd t},
\end{align*}
and a corresponding experiment
\begin{gather*}
    \pi^*(q,t)=
    \begin{cases}
        1 & \text{for all } q, t  \text{ such that } \beta(q) > -\phi_c^+(t) \\
        D & \text{for all } q, t  \text{ such that }  \beta(q) = -\phi_c^+(t) \\
        0 & \text{otherwise}
    \end{cases}.
\end{gather*}

This experiment gives rise to a threshold mechanism by using the payment function defined in Theorem \ref{thm:opt-scheme}.   Notably, when $\beta(q) =-\phi^+_{c}(t) $ doesn't have point mass at this $c$ point,  $D$ will be 0 due to continuity and this degenerates to the threshold experiment for Case 3 in Theorem \ref{thm:opt-scheme}. The feasibility and the optimality of the above mechanisms follow from the same   argument  in the proof of Lemma \ref{lem:case_3_feasible} and Proposition \ref{lem:case_3}, essentially because the boundary case of $ \beta(q) = -\phi_c^+(t)$ will not affect revenue. We   omit details here. 
\clearpage

\end{document}